\documentclass[pra, twocolumn,amsfonts,amsmath,amssymb,eufrak]{revtex4-1}

\usepackage{epsfig}
\usepackage{color}
\usepackage{bm}
\usepackage{hyperref}
\usepackage[dvipsnames]{xcolor}
\hypersetup{
    bookmarksnumbered=true, 
    unicode=false, 
    pdfstartview={FitH}, 
    pdftitle={}, 
    pdfauthor={}, 
    pdfsubject={}, 
    pdfcreator={}, 
    pdfproducer={}, 
    pdfkeywords={}, 
    pdfnewwindow=true, 
    colorlinks=true, 
    linkcolor=NavyBlue, 
    citecolor=NavyBlue, 
    filecolor=NavyBlue, 
    urlcolor=NavyBlue 
}

\usepackage{braket}
\usepackage{comment}
\usepackage{graphicx}
\usepackage{amsthm}
\usepackage{ragged2e}
\usepackage{physics}
\usepackage[caption=false]{subfig}

\newcommand{\figin}[4]
{\begin{figure}[tb]
\includegraphics[width= #1]{#2.pdf}
\caption{#3}
\label{f:#4}
\end{figure}}

\newcommand{\todayd}{\the\year/\the\month/\the\day}

\newcommand{\bib}{\bibitem}

\newcommand{\bel}{\begin{easylist}}
\newcommand{\eel}{\end{easylist}}

\newcommand{\fref}[1]{Fig.~\ref{f:#1}}

\def \({\left(}
\def \){\right)}
\def \[{\left[}
\def \]{\right]}

\newcommand{\sumtwo}[2]%
{\mathop{\sum_{#1}}_{#2}}
\newcommand{\sumthree}[3]%
{\mathop{\mathop{\sum_{#1}}_{#2}}_{#3}}
\newcommand{\sumfour}[4]%
{\mathop{\mathop{\mathop{\sum_{#1}}_{#2}}_{#3}}_{#4}} 
\newcommand{\prodtwo}[2]%
{\mathop{\prod_{#1}}_{#2}}
\newcommand{\mintwo}[2]%
{\mathop{\min_{#1}}_{#2}}
\newcommand{\maxtwo}[2]%
{\mathop{\max_{#1}}_{#2}}
\newcommand{\maxthree}[3]%
{\mathop{\mathop{\max_{#1}}_{#2}}_{#3}}
\newcommand{\limtwo}[2]%
{\mathop{\lim_{#1}}_{#2}}
\newcommand{\suptwo}[2]%
{\mathop{\sup_{#1}}_{#2}}
\newcommand{\supthree}[3]%
{\mathop{\mathop{\sup_{#1}}_{#2}}_{#3}}
\newcommand{\supfour}[4]%
{\mathop{\mathop{\mathop{\sup_{#1}}_{#2}}_{#3}}_{#4}} 
\newcommand{\inftwo}[2]%
{\mathop{\inf_{#1}}_{#2}}
\newcommand{\infthree}[3]%
{\mathop{\mathop{\inf_{#1}}_{#2}}_{#3}}
\newcommand{\inffour}[4]%
{\mathop{\mathop{\mathop{\inf_{#1}}_{#2}}_{#3}}_{#4}} 

\newcommand{\para}[1]{{\em #1}\/.---}

\newtheorem{thm}{Theorem}
\newtheorem{lm}[thm]{Lemma}
\newtheorem{pro}[thm]{Proposition}

\newcommand{\bthm}[1]{\begin{thm} #1 \end{thm}}

\theoremstyle{definition}
\newtheorem{dfn}{Definition}

\hypersetup{
  pdftitle={Optimal asymptotic pure-state conversion under thermal operations},
  pdfauthor={Tatsuya Yamanouchi and Naoto Shiraishi}
}

\theoremstyle{plain}
\newtheorem{theorem}{Theorem}
\newtheorem{proposition}[theorem]{Proposition}
\newtheorem{lemma}[theorem]{Lemma}

\theoremstyle{definition}

\theoremstyle{remark}
\newtheorem{remark}{Remark}

\begin{document}

\title{Optimal asymptotic pure-state conversion under thermal operations}

\author{Tatsuya Yamanouchi}
\email{yamanouchitatsuya@gmail.com}
\author{Naoto Shiraishi} 
\email{shiraishi@phys.c.u-tokyo.ac.jp}
\affiliation{Department of Basic Science, The University of Tokyo, 3-8-1 Komaba, Meguro-ku, Tokyo 153-8902, Japan}

\begin{abstract}
Quantum thermodynamics seeks to characterize possible state transformations under thermodynamic constraints in microscopic quantum systems. In the fully quantum regime, however, quantum coherence imposes various restrictions on state conversions, and a complete characterization of state convertibility has remained open. Here we resolve this problem for asymptotic conversions between pure states. Specifically, we show that asymptotic convertibility is fully determined by the free-energy and energy-variance ratios. Our result reveals how genuinely quantum effects quantitatively constrain state conversions in quantum thermodynamics. 
\end{abstract}

\maketitle

Thermodynamics and quantum mechanics are two cornerstones of modern physics, describing phenomena at distinct scales. Understanding how thermodynamic principles manifest at microscopic scales, where genuinely quantum effects become relevant, is therefore a fundamental problem at the interface of the two theories and has attracted broad interest in quantum information theory and statistical physics. A central question is to determine which transformations between microscopic quantum states are possible in a thermal environment and, in turn, which properties of a quantum state constitute thermodynamic resources. The resource theory of thermodynamics provides an operational framework for addressing this question, where we ask which state conversions are possible under thermodynamically allowed operations~\cite{Gour-review, Lostaglio-review, Sagawa-review}. A standard and operationally motivated class of such operations is the class of thermal operations (TO), in which a system interacts with a heat bath in the Gibbs state through an energy-conserving unitary \cite{Janzing-2000, Horodecki-2013, Brandao-2013}.

Thermal operations are mainly subject to two different constraints. One is the Gibbs-preserving condition that the Gibbs state of the input system is mapped to the Gibbs state of the output system, imposing restrictions on athermality \cite{Horodecki-2013, Brandao-2013}. Quantum channels satisfying this condition are called Gibbs-preserving operations (GPO). In the classical regime, this property fully characterizes state convertibility under TO, and state conversion conditions under TO and GPO coincide~\cite{Horodecki-2013, Shiraishi-2020}. In the quantum regime, however, some state conversions possible under GPO are impossible under TO~\cite{Faist-2015, Tajima-2025}. This limitation originates from another property of TO, time-translation covariance, which arises from the law of energy conservation and imposes genuinely quantum restrictions on the manipulation of coherent superpositions of energy eigenstates \cite{Lostaglio-2015, Cwiklinski-2015}. The central challenge is to understand how these two types of constraints jointly determine possible state conversions in the quantum regime. In this work, we tackle this problem in asymptotic conversions, where independent and identically distributed (i.i.d.) copies of one quantum state are converted into i.i.d. copies of another.

Asymptotic state convertibility under one of the two constraints is well understood. Under Gibbs-preserving operations, state convertibility is characterized solely by the free energy defined through quantum relative entropy with respect to the Gibbs state \cite{Matsumoto-2010, Buscemi-2019}. On the other hand, covariant operations (CO), those satisfying only the time-translation covariance, have been studied in the resource theory of asymmetry \cite{Marvian-2013, Marvian-2014, Marvian-2020, Marvian-2022, Yamaguchi-2026-pure, Yamaguchi-2026-mixed}, and in particular, asymptotic state convertibility between pure states is characterized by the energy variance \cite{Marvian-2022, Yamaguchi-2026-pure}.

Having understood state convertibility under GPO and CO separately, one might expect that the constraints imposed by TO could be obtained simply by combining the two corresponding convertibility conditions. This expectation, however, is not justified in general. To see why, we introduce a relaxation of thermal operations called covariant Gibbs-preserving operations (CGPO), defined as the set of quantum channels that are simultaneously Gibbs-preserving and time-translation covariant \cite{Cwiklinski-2015}. (See \fref{operation-class}.)
\figin{0.6\columnwidth}{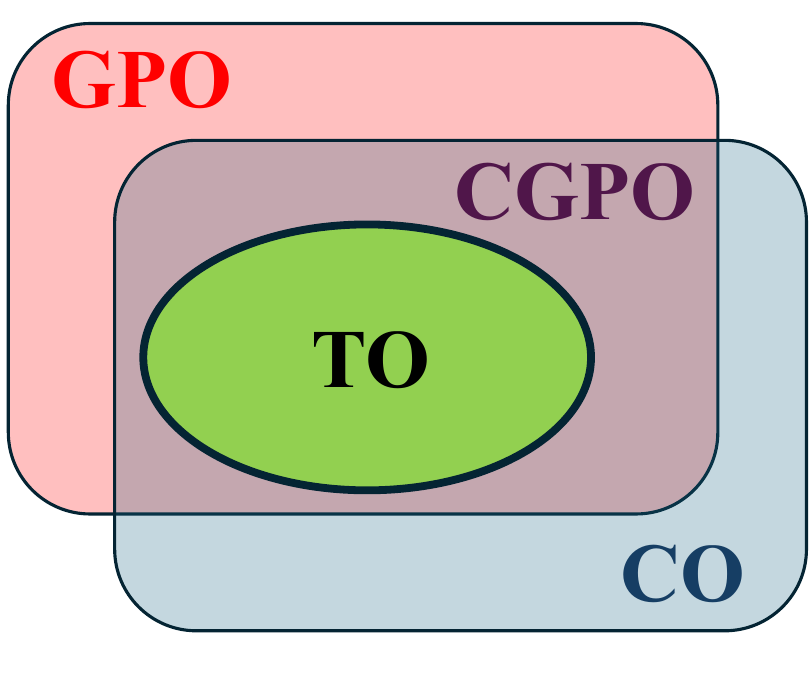}{Relations among the relevant classes of operations. The set of covariant Gibbs-preserving operations (CGPO) is the intersection of the set of Gibbs-preserving operations (GPO) and the set of covariant operations (CO), and thermal operations (TO) form a strict subset of CGPO.}{operation-class} 
Indeed, in the single-shot regime, there exist transformations that are separately possible under GPO and CO but are impossible under  CGPO, meaning that the simultaneous imposition of the two constraints results in additional restrictions on state conversions~\cite{Gour-2018}. Moreover, the set of CGPO is strictly larger than the set of thermal operations~\cite{Ding-2021}. Consequently, state convertibility under TO cannot, in general, be inferred simply by combining the convertibility conditions for GPO and CO.

Despite being one of the most fundamental and physically motivated classes of operations in quantum thermodynamics, the conversion power of TO has not been fully characterized in the genuinely quantum regime. In the asymptotic setting, when the target state is energy-incoherent, asymptotic convertibility under thermal operations is fully determined by the free energy even when the initial state has energetic coherence~\cite{Brandao-2013, Lipka-Bartosik-2024}. For conversions involving coherence in both the input and output states, however, a complete characterization of asymptotic convertibility has remained open. Previous studies on this problem have shown that, by supplementing the operations with an asymptotically negligible amount of additional resources, such as coherence or energy, state conversions are fully governed by the free energy~\cite{Brandao-2013,Sagawa-2021, Faist-2019, Gour-2022}. While these formulations successfully derived necessary and sufficient conditions for asymptotic state conversions, they modify the original conversion problem by introducing additional resources and therefore do not characterize the intrinsic conversion power of thermal operations themselves. A related complete characterization has been obtained in the marginal asymptotic regime, where only the reduced state of each output copy is required to approach the target state. In this setting, state convertibility is characterized essentially by the free energy~\cite{Shiraishi-2025, Shiraishi-Takagi-2025}. However, this notion substantially relaxes the original i.i.d. conversion problem, and hence does not determine the conversion power of thermal operations in the standard asymptotic setting considered here.

Against this backdrop, we study asymptotic pure-state convertibility under thermal operations themselves in the standard i.i.d. asymptotic regime, and show that state convertibility is completely determined by the two constraints associated with GPO and CO. Specifically, the optimal conversion rate is exactly the minimum of the free-energy ratio and the energy-variance ratio. Our result reveals that, in sharp contrast to the single-shot setting, no additional constraints exist in asymptotic conversions between pure states. To the best of our knowledge, this provides the first complete characterization of asymptotic pure-state convertibility under thermal operations allowing energetic coherence in both the input and target states, without introducing additional resources.


\para{Setup and main result}
We consider asymptotic state conversion under thermal operations. We first  define thermal operations. For a fixed inverse temperature $\beta>0$, we denote the Gibbs state of system $X$ with Hamiltonian $H_X$ by $\gamma_X := e^{-\beta H_X} / \Tr[e^{-\beta H_X}]$. A quantum channel $\mathcal{E}:\mathcal{B}(S_1)\to \mathcal{B}(S_2)$ is called a thermal operation if and only if there exist an auxiliary system $E_1$ and an energy-conserving unitary $U$ (i.e., a unitary that commutes with the total Hamiltonian $H_{S_1} + H_{E_1}$) on $S_1 \otimes E_1$ such that
\begin{align}
    \mathcal{E}(\,\cdot\,) = \Tr_{E_2}\!\left[U\left(\,\cdot\,\otimes\gamma_{E_1}\right)U^\dagger\right].
\end{align}
Here, the total Hilbert space and the Hamiltonian of the input and output systems are the same: $S_1 \otimes E_1=S_2 \otimes E_2$ and $H_{S_1}+H_{E_1} = H_{S_2}+H_{E_2}$.

We consider a finite-dimensional quantum system $S$ with Hamiltonian $H_S$ whose energy-level spacings are integer multiples of some common constant. In the following, we consider $n$ copies of system $S$ with Hamiltonian $H_S^{(n)}:=\sum_{k=1}^{n} I^{\otimes (k-1)}\otimes H_S \otimes I^{\otimes (n-k)}$. A rate $R\ge 0$ is said to be achievable under thermal operations if there exist sequences of integers $\{m_n\}_{n=1}^\infty$ and thermal operations $\{\mathcal{E}_n:\mathcal{B}(S^{\otimes n})\to \mathcal{B}(S^{\otimes m_n})\}_{n=1}^\infty$ such that $d_\mathrm{tr}(\mathcal{E}_n(\psi^{\otimes n}), \phi^{\otimes m_n})\to 0$ and $m_n/n\to R$ as $n\to\infty$, where $d_\mathrm{tr}(\rho,\sigma):=\|\rho-\sigma\|_1/2$ is the trace distance between quantum states. The optimal conversion rate $R_\mathrm{TO}(\psi\to\phi)$ is defined as the supremum of all rates achievable under thermal operations. 

Thermal operations satisfy two important properties, namely, the Gibbs-preserving condition: $\mathcal{E}(\gamma_{S_1})=\gamma_{S_2}$, and the time-translation covariance: $\mathcal{E}\circ\mathcal{U}_t^{S_1}=\mathcal{U}_t^{S_2}\circ\mathcal{E}$ for all $t\in\mathbb R$, where $\mathcal{U}^X_t(\,\cdot\,)=e^{-iH_X t}(\,\cdot\,) e^{iH_X t}$ is the time-evolution map generated by the Hamiltonian $H_X$. The Gibbs-preserving condition imposes restrictions on the manipulation of athermality, whereas the time-translation covariance constrains coherent superpositions of different energy eigenstates. Intuitively, the latter property implies that thermal operations cannot increase the amount of energetic coherence, nor can they create coherence from states without it. To formulate this property precisely, we introduce the time-translation symmetry subgroup~\cite{Marvian-2013}, which is defined as the set of all periods of a state  by
\begin{align}
    \mathrm{Sym}(\rho)
    :=
    \left\{t\in\mathbb{R}:\mathcal{U}_t^S(\rho)=\rho\right\}.
\end{align}
Using this quantity, the restriction that coherence cannot be generated from scratch can be expressed as $\mathrm{Sym}(\rho)\subseteq \mathrm{Sym}(\sigma)$, which is a necessary condition for converting $\rho$ into $\sigma$ under thermal operations. We refer to this as the symmetry condition.

In the following, the relative entropy and the energy variance are defined as $D(\rho\|\sigma) := \Tr [\rho\log \rho - \rho \log \sigma]$ and $V_{H_S}(\rho) := \Tr [H_S^2 \rho] - (\Tr [H_S \rho])^2$. Our main result is stated as follows.
\bthm{\label{them:pure-TO}
Let $S$ be a finite-dimensional quantum system whose energy-level spacings are integer multiples of a common constant, and let $\psi$ and $\phi$ be pure states on $S$. If $\mathrm{Sym}(\psi) \subseteq \mathrm{Sym}(\phi)$, then the optimal conversion rate is given by
\begin{align}\label{eq:main-rate}
    R_\mathrm{TO}(\psi\to\phi) 
    =
    \min\left\{
    \dfrac{D(\psi\|\gamma_S)}{D(\phi\|\gamma_S)},
    \dfrac{V_{H_S}(\psi)}{V_{H_S}(\phi)}
    \right\},
\end{align}
otherwise, $R_\mathrm{TO}(\psi\to\phi)=0$.}

In the case of $V_{H_S}(\phi)=0$, we adopt the convention $V_{H_S}(\psi)/V_{H_S}(\phi)=\infty$ including when $V_{H_S}(\psi)= 0$.

Theorem~\ref{them:pure-TO} provides a complete characterization of asymptotic pure-state conversion under thermal operations. The optimal conversion rate under Gibbs-preserving operations is given by $D(\psi\|\gamma_S)/D(\phi\|\gamma_S)$~\cite{Buscemi-2019}, whereas the optimal conversion rate under covariant operations is determined by $V_{H_S}(\psi)/V_{H_S}(\phi)$, together with the symmetry condition~\cite{Marvian-2022, Yamaguchi-2026-pure}. Thus, one immediately obtains an upper bound on $R_\mathrm{TO}(\psi\to\phi)$ given by the minimum of the free-energy ratio and the energy-variance ratio. It is, however, highly nontrivial whether this upper bound can be saturated under the restrictive class of thermal operations. Our theorem shows that any rates below the two bounds are indeed achievable, and the optimal rate under thermal operations is simply obtained by combining these two constraints.

The above theorem also holds without modification for CGPO. Therefore, in sharp contrast to the single-shot setting, where TO and CGPO exhibit different state-conversion capabilities~\cite{Ding-2021}, they have the same state conversion power for asymptotic transformations between pure states. A similar collapse has recently been established in the correlated-catalytic setting, where the conversion conditions for TO and CGPO reduce to those obtained by combining the corresponding conditions for GPO and CO~\cite{Shiraishi-Sagawa-2021, Shiraishi-Takagi-2024, Kondra-2024, Shiraishi-2025, Shiraishi-Takagi-2025}. Taken together, our result and these correlated-catalytic results suggest that the requirement of microscopic implementation distinguishing TO from CGPO may be sufficiently mild that its effect disappears under different forms of regularization.

\para{Physical implications}
To illustrate the physical implications of the above theorem, we consider a qubit system with Hamiltonian $H_S = \varepsilon \ketbra{1}{1}\, (\varepsilon>0)$, and the asymptotic conversion from
\begin{align}
    \ket{\psi} = \frac{1}{\sqrt{2}}\ket{0}+\frac{1}{\sqrt{2}}\ket{1}
\end{align}
to
\begin{align}
    \ket{\phi} = \sqrt{\frac{2}{3}}\ket{0}+{\frac{1}{\sqrt{3}}}\ket{1}.
\end{align}
The free energies are given by $D(\psi\|\gamma_S) = \log Z + \beta\varepsilon/2$ and $D(\phi\|\gamma_S) = \log Z + \beta\varepsilon/3$, where $Z:=1+e^{-\beta \varepsilon}$ is the partition function, whereas the energy-variance ratio is $V_{H_S}(\psi)/V_{H_S}(\phi)= 9/8$, which is independent of the temperature. The two constraints  therefore respond differently to a change in temperature. In the high-temperature limit  $\beta\varepsilon\to 0$, the free-energy ratio approaches unity so that the conversion rate is limited by the free-energy constraint. By contrast, in the low-temperature limit $\beta\varepsilon\to\infty$, the free-energy ratio approaches $3/2$, and the energy variance constraint becomes the bottleneck, fixing the optimal rate at $9/8$. Thus, even for a fixed pair of pure states, the limiting constraint on asymptotic state conversion can change with temperature. In addition, the round-trip rate satisfies $R_\mathrm{TO}(\psi\to\phi)R_\mathrm{TO}(\phi\to\psi)<1$ except for a specific temperature at which $D(\psi\|\gamma_S)/D(\phi\|\gamma_S) = V_{H_S}(\psi)/V_{H_S}(\phi)$, and hence the conversion is generally irreversible. Thus, unlike in the asymptotic resource theory of quasiclassical states, where the free energy alone determines state convertibility and the theory is reversible, energetic coherence introduces an additional independent constraint that generically destroys reversibility.


\para{Sketch of the proof of Theorem~\ref{them:pure-TO}}
Here we give an overview of the proof of our theorem. A complete proof is provided in the Supplemental Material~\cite{SM}. Since the converse bound follows directly from the Gibbs-preserving condition and the time-translation covariance, our main task is to establish achievability: for two pure states satisfying $\mathrm{Sym}(\psi)\subseteq \mathrm{Sym}(\phi)$, we  construct a sequence of thermal operations for any conversion rate $R$ below the claimed optimal rate. 

To construct the desired conversion protocol, we use the fact that an approximate transformation between pure states by an energy-conserving unitary is possible if the input and output energy distributions are close in total variation distance~\cite{Marvian-2013, Marvian-2022}. Our strategy is therefore to compensate for the mismatch between the energy distributions of $\ket{\psi}^{\otimes n}$ and $\ket{\phi}^{\otimes m_n}$ by introducing a suitable auxiliary system in the Gibbs state. The key observation is that, for i.i.d. copies of a pure state, the energy distribution is asymptotically governed only by its mean and variance~\cite{Marvian-2022}. By appropriately choosing the auxiliary system so as to match these two quantities between the input and output, we make the energy distributions of the input and output systems asymptotically coincide in total variation distance. Because the auxiliary Gibbs state is mixed, we need to ensure the above energy-distribution matching simultaneously for all typical energy eigenstates of the Gibbs state. We then construct a single energy-conserving unitary that maps every typical input state to the corresponding output state. Since the atypical components have vanishing total weight, this yields the desired thermal operation with asymptotically vanishing error.


In the following, we explain the details of the protocol. Here we assume that both $\psi$ and $\phi$ have energetic coherence, i.e., not energy eigenstates, and have the same time-translation symmetry: $\mathrm{Sym}(\psi)= \mathrm{Sym}(\phi)\subsetneq \mathbb{R}$. Conversions between energy eigenstates can be treated by essentially the same construction with minor simplifications. The case in which the symmetry subgroup of $\psi$ is strictly contained in that of $\phi$ can be reduced to the case in which the two symmetry groups coincide. 

We construct a protocol that converts $\psi^{\otimes n}$ to $\phi^{\otimes \lfloor Rn \rfloor}$. We introduce an auxiliary system composed of $B$ and $B'$. The first bath $B$, consisting of many copies of the original system $S$, is used to absorb the difference in mean energy between $\psi^{\otimes n}$ and $\phi^{\otimes \lfloor Rn \rfloor}$. We set $B = S^{\otimes \ell}$ where $\ell= \lfloor n^{3/2}\rfloor$. Then its Gibbs state $\gamma_B$ is highly concentrated on a typical eigenspace, denoted by $\mathcal{T}_\ell$, with fluctuation in the occupation number of  $O(\sqrt{\ell})$. In the asymptotic limit $n\to\infty$, the weight of $\gamma_B$ outside $\mathcal{T}_\ell$ converges to zero. Note that the choice $\ell= \lfloor n^{3/2}\rfloor$ ensures that $O(\sqrt{\ell}) = o(n)$ and hence the fluctuation is asymptotically negligible. The second bath is a specially designed system whose role is to absorb the difference in energy variance between $\psi^{\otimes n}$ and $\phi^{\otimes \lfloor Rn \rfloor}$. We fix integers $\Delta_n:=\lfloor n^{3/4}\rfloor$ and $N_n:=\lfloor \sqrt{n}\log n\rfloor$. We choose $B'$ as a system with a nondegenerate ground state with zero energy and $2N_n+1$ excited states with energies around $E_*\Delta_n$ with spacing $E_*$, where $E_*$ denotes the energy gap corresponding to the common minimal period of $\psi$ and $\phi$. Since $\Delta_n\gg N_n$, the Gibbs state of $B'$ is asymptotically concentrated on the ground state:
$\gamma_{B'} \simeq \ketbra{g}{g}$.

With the above choice of auxiliary systems, we fix a typical state $\ket{\lambda}\in\mathcal{T}_\ell$ of $B$, and compare the energy distributions of the input and output states
\begin{align}
    \ket{\Psi_\lambda}:= \ket{\psi}^{\otimes n}\ket{\lambda}\ket{g}
    ,\quad
    \ket{\Phi_\lambda}:= \ket{\phi}^{\otimes \lfloor Rn\rfloor}\ket{t(\lambda)}\ket{\eta_n}.
\end{align}
Here, $\ket{t(\lambda)}$ and $\ket{\eta_n}$ are output states of $\widetilde{B} = S^{\otimes (n+\ell -\lfloor Rn\rfloor)}$ and $B'$, respectively. The energy distribution of i.i.d. copies of a pure state is asymptotically approximated by a translated Poisson distribution characterized by its mean and variance~\cite{Marvian-2022}. Therefore, to show that the energy distributions of $\ket{\Psi_\lambda}$ and $\ket{\Phi_\lambda}$ are close, it suffices to show that we can set $\ket{t(\lambda)}$ and $\ket{\eta_n}$ such that the mean energies and the energy variances of $\ket{\Psi_\lambda}$ and $\ket{\Phi_\lambda}$ are close. Here, $\ket{t(\lambda)}$ is an energy eigenstate of $\widetilde{B}$, which compensates for the change in mean energy from $\ket{\psi}^{\otimes n}$ to $\ket{\phi}^{\otimes \lfloor Rn\rfloor}$, and $\ket{\eta_n}$ is a superposition over $2N_n+1$ excited states of $B'$, whose energy variance is close to $V_{\eta_n}:=nV_{H_S}(\psi)-\lfloor Rn\rfloor V_{H_S}(\phi)$. The energy-variance bound $R < V_{H_S}(\psi)/V_{H_S}(\phi)$ ensures that $V_{\eta_n}>0$ and the above construction is indeed possible. In this way, we show that, for each typical state $\ket{\lambda}\in\mathcal{T}_\ell$, the total variation distance between the energy distributions of $\ket{\Psi_\lambda}$ and $\ket{\Phi_\lambda}$ converges to zero as $n\to\infty$.
\figin{1\columnwidth}{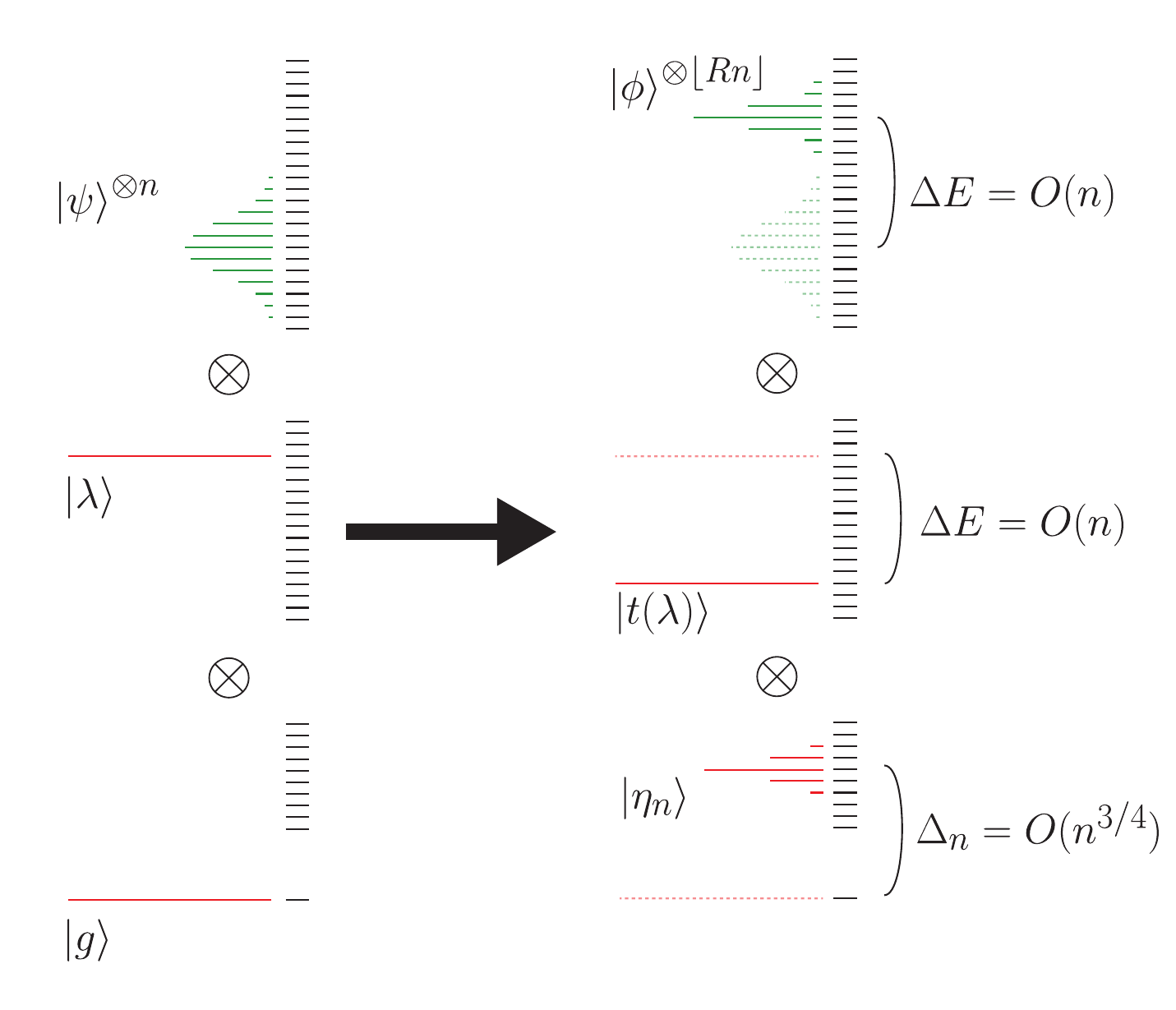}{Schematic illustration of the energy-distribution matching for each typical bath state $\ket{\lambda}\in\mathcal{T}_\ell$. The energy eigenstate $\ket{t(\lambda)}$ is chosen so as to compensate for the $O(n)$ difference in the mean energies of the input and output states, and the auxiliary state $\ket{\eta_n}$ has an energy variance of $O(n)$ that adjusts the difference in energy variance between the input and output states. The mean energy of $\ket{\eta_n}$ is of order $O(\Delta_n)=o(n)$ and is therefore negligible in the asymptotic rate.}{protocol} 

Now we define a unitary $V$ that converts $\ket{\Psi_\lambda}$ to $\ket{\Phi_\lambda}$ for all typical states $\ket{\lambda}\in\mathcal{T}_\ell$ of bath $B$. The free-energy bound $R < D(\psi\|\gamma_S)/D(\phi\|\gamma_S)$ ensures that there are sufficiently many states $\ket{t(\lambda)}$ satisfying the requirements in the above energy-distribution matching, so that the assignment $\ket{\lambda}\mapsto \ket{t(\lambda)}$ can be chosen to be injective. Consequently, within each energy sector of total energy $E$, we can define a unitary $V_E$ that maps the normalized energy-$E$ component of each $\ket{\Psi_\lambda}$ to the corresponding normalized energy-$E$ component of $\ket{\Phi_\lambda}$. Taking the direct sum over all energy sectors $V:=\bigoplus_E V_E$,
we obtain an energy-conserving unitary that realizes the transformation $\ket{\Psi_\lambda}\to\ket{\Phi_\lambda}$ for every typical $\lambda$, with an error controlled by the total variation distance between their energy distributions. Since this distance vanishes in the asymptotic limit, the energy-conserving unitary $V$ implements the desired transformation for all typical bath states with asymptotically vanishing error.

Combining the above construction for all typical bath states and noting that the atypical contribution has vanishing weight and $\gamma_{B'}\simeq \ketbra{g}{g}$, we obtain a thermal operation that converts $\psi^{\otimes n}$ to $\phi^{\otimes \lfloor Rn \rfloor}$ with asymptotically vanishing error. This completes the construction of the desired conversion protocol.

\para{Conclusion}
We have established a complete characterization of i.i.d. asymptotic conversion between pure states under thermal operations, including states with energetic coherence. Our result shows that, once the symmetry condition is satisfied, asymptotic conversions between pure states are fully governed by two resources: free energy and energy variance. Thus, the microscopic implementability required in thermal operations introduces no further asymptotic restriction beyond those imposed separately by GPO and CO.

Our result also clarifies the role of coherence assistance employed in previous studies of asymptotic state conversion. Such assistance was known to enable transformations that are otherwise impossible under thermal operations, but the thermodynamic resource supplied by the auxiliary coherence was not transparent. Our characterization identifies the coherence-related restrictions compensated for by such assistance as precisely those captured by the energy-variance and symmetry conditions.

We comment on our assumption on the Hamiltonian, namely, that all energy-level spacings are integer multiples of some constant. This assumption excludes Hamiltonians with irrational ratios between energy gaps. For general Hamiltonians, one can instead decompose the energy gaps into rationally independent components and analyze the corresponding sectors separately. We expect that an analogous characterization can be established in this more general setting.

Extending our result to general mixed states appears substantially more challenging. Our achievability proof makes essential use of the purity of the input and output states, in particular in reducing the problem to matching their energy distributions. For mixed states, however, the energy distributions alone are insufficient to guarantee the existence of such a unitary, and a simple and elementary protocol can no longer be expected. Consequently, the present approach based on matching energy distributions does not directly extend to mixed states. This more general setting will be addressed in separate work using a different approach.

{\it Acknowledgments.}---
The authors thank Ryuji Takagi, Kaito Watanabe, and Yosuke Mitsuhashi for fruitful discussions. OpenAI ChatGPT (GPT-5.4, GPT-5.5, and GPT-5.6 Sol) was used to verify and simplify the construction of the conversion protocol. All proofs and conclusions presented in this work were independently verified by the authors. This work was supported by Grant-in-Aid for Early-Career Scientists 26K17048, Grant-in-Aid for Transformative Research Areas (B) 26K00021, JST ERATO Grant Number JPMJER2302, and World-Leading Innovative Graduate Study Program for Advanced Basic Science Course at the University of Tokyo.

\makeatletter
\let\auto@bib@innerbib\@empty
\makeatother

\begingroup
\renewcommand{\addcontentsline}[3]{}

\endgroup

\clearpage

\onecolumngrid

\pagestyle{plain}
\setcounter{equation}{0}
\renewcommand{\theequation}{S.\arabic{equation}}

\setcounter{theorem}{0}
\renewcommand{\thetheorem}{S.\arabic{theorem}}

\setcounter{figure}{0}
\renewcommand{\thefigure}{S.\arabic{figure}}

\setcounter{table}{0}
\renewcommand{\thetable}{S.\arabic{table}}

\begin{center}

{\large\bfseries
Supplemental Material for\\
``Optimal asymptotic pure-state conversion under thermal operations''
\par}

\vspace{0.3cm}

Tatsuya Yamanouchi and Naoto Shiraishi

\vspace{0.15cm}

{\itshape Department of Basic Science, The University of Tokyo}

\end{center}

\vspace{0.4cm}

In this Supplemental Material, we prove our main theorem. The contents are organized as follows. In Section~\ref{supp:sec:preliminaries}, we introduce the basic concepts needed to state our result and review known results used in the proof. In Section~\ref{supp:sec:main-result}, we state our main result. The proof of the main theorem is given in Sections~\ref{supp:sec:converse}--\ref{supp:sec:technical-lemmas}. In Section~\ref{supp:sec:converse}, we prove the converse part, and in Section~\ref{supp:sec:achievability}, we prove the achievability part. In Section~\ref{supp:sec:technical-lemmas}, we provide proofs of technical lemmas used in Sections~\ref{supp:sec:converse} and~\ref{supp:sec:achievability}.

\tableofcontents

\section{Preliminaries}\label{supp:sec:preliminaries}

In this section, we introduce basic concepts which are necessary to describe our result and review known results used in the proof. Throughout this work, all Hilbert spaces are assumed to be finite-dimensional and all logarithms are natural logarithms. We use units in which $k_B = \hbar = 1$.

For a quantum system $X$, we denote its Hilbert space by $\mathcal H_X$ and its Hamiltonian by $H_X$. We fix an inverse temperature $\beta>0$, and write the Gibbs state of $X$ as
\begin{align}
    \gamma_X
    :=
    \frac{e^{-\beta H_X}}{\Tr[ e^{-\beta H_X}]} .
\end{align}
The set of density operators acting on $\mathcal{H}_X$ is denoted by $\mathcal{D}(\mathcal{H}_X)$. For a pure state $\ket{\psi}\in \mathcal{H}_X$, we use the notation $\psi := \ketbra{\psi}{\psi} \in \mathcal{D}(\mathcal{H}_X)$.

\subsection{Thermal operations}
\subsubsection{Definition of thermal operations}
Thermal operations offer a standard and operationally motivated framework for investigating state convertibility of quantum systems in contact with a thermal bath~\cite{supp:Horodecki-2013,supp:Brandao-2013}. We define a thermal operation as follows: a completely positive and trace-preserving (CPTP) map $\mathcal{E}:\mathcal{B}(\mathcal{H}_S)\to\mathcal{B}(\mathcal{H}_{S'})$ is a thermal operation if and only if there exist an auxiliary system $E$ and an energy-conserving unitary $U$ on $\mathcal{H}_S\otimes \mathcal{H}_E$ (i.e., a unitary that commutes with the total Hamiltonian $H_\mathrm{tot}$) such that
\begin{align}
    \mathcal{E}(\, \cdot\,) = \Tr_{E'}\!\left[U(\ \cdot\ \otimes \gamma_E)U^\dagger\right].
\end{align}
Here, $E'$ is a subsystem to be discarded, satisfying
\begin{align}
    \mathcal{H}_S \otimes\mathcal{H}_E = \mathcal{H}_{S'} \otimes\mathcal{H}_{E'}
\end{align}
and
\begin{align}
    H_\mathrm{tot} = H_S\otimes I_E+I_S\otimes H_E=H_{S'}\otimes I_{E'}+I_{S'}\otimes H_{E'}.
\end{align}
In the following, we often omit identity operators and write, for example, $H_\mathrm{tot} = H_S + H_E$.

\subsubsection{Asymptotic state conversion under thermal operations}\label{supp:subsubsec:asymptotic-TO}

We consider asymptotic state conversion under thermal operations. Let $\rho\in \mathcal{D}(\mathcal{H}_S)$ be the input state and $\sigma\in \mathcal{D}(\mathcal{H}_S)$ be the target state. For each $n$, we consider a transformation between independent and identically distributed (i.i.d.) copies of quantum states
\begin{align}
    \rho^{\otimes n}\longrightarrow\sigma^{\otimes m_n}
\end{align}
implemented by a thermal operation with a vanishing error as $n\to\infty$. The Hilbert space associated with the input system $S^{\otimes n}$ is $\mathcal{H}_S^{\otimes n}$, and the Hamiltonian of $S^{\otimes n}$ is 
\begin{align}
    H_S^{(n)} := \sum_{k=1}^n I^{\otimes k-1}\otimes H_S \otimes I^{\otimes (n-k)}.
\end{align}
Similarly, the Hilbert space and Hamiltonian of the output system $S^{\otimes m_n}$ are $\mathcal{H}_S^{\otimes m_n}$ and $H_S^{(m_n)}$.

A conversion rate $R\geq 0$ is said to be achievable under thermal operations if there exist sequences of thermal operations $\{\mathcal E_n\}_{n=1}^{\infty}$ and integers $\{m_n\}_{n=1}^{\infty}$ such that
\begin{align}
    \lim_{n\to\infty}d_\mathrm{tr}\left(\mathcal{E}_n (\rho^{\otimes n}), \sigma^{\otimes m_n}\right)=0
\end{align}
and
\begin{align}
    \lim_{n\to\infty}\frac{m_n}{n} = R.
\end{align}
Here, $d_\mathrm{tr}(\rho,\sigma)$ is the trace distance defined by
\begin{align}
    d_\mathrm{tr}(\rho,\sigma) := \frac{1}{2}\|\rho - \sigma\|_1,
\end{align}
where $\|\,\cdot\,\|_1$ denotes the trace norm, defined for an operator $A$ by 
\begin{align}
    \|A\|_1 := \Tr \left[ \sqrt{A^\dagger A}\right].
\end{align}
The optimal asymptotic conversion rate is defined by
\begin{align}
    R_{\mathrm{TO}}(\rho\to\sigma):=
    \sup\left\{R:R\text{ is achievable under thermal operations}\right\}.
\end{align}

\subsection{Related classes of operations}

Thermal operations are subject to two constraints: Gibbs-preservation and time-translation covariance \cite{supp:Lostaglio-2019}. The simultaneous imposition of these constraints makes thermal operations difficult to analyze. It is therefore convenient to consider two larger classes of operations, each defined by imposing only one of the two constraints. Here we introduce these two classes and review existing results on asymptotic convertibility under them.

\subsubsection{Gibbs-preserving operations}

Thermal operations preserve the Gibbs state. More precisely, if $\mathcal{E}:\mathcal{B}(\mathcal{H}_{S_1})\to\mathcal{B}(\mathcal{H}_{S_2})$ is a thermal operation, then it maps the Gibbs state of the input system to that of the output system:
\begin{align}
    \mathcal{E}(\gamma_{S_1}) = \gamma_{S_2}.
\end{align}
Motivated by this property, we define Gibbs-preserving operations as follows. A CPTP map $\mathcal{E}:\mathcal{B}(\mathcal{H}_{S_1})\to\mathcal{B}(\mathcal{H}_{S_2})$ is a Gibbs-preserving operation if and only if  $\mathcal{E}(\gamma_{S_1}) = \gamma_{S_2}$. By definition, every thermal operation is a Gibbs-preserving operation. This inclusion is known to be strict: there exist Gibbs-preserving operations that are not thermal operations~\cite{supp:Faist-2015}.

Asymptotic state conversion under Gibbs-preserving operations is defined analogously to that under thermal operations in Section~\ref{supp:subsubsec:asymptotic-TO}. The optimal conversion rate under Gibbs-preserving operations, denoted by $R_\mathrm{GPO}(\rho\to\sigma)$, is defined as the supremum of all conversion rates achievable under Gibbs-preserving operations. Asymptotic conversions under Gibbs-preserving operations are characterized by the nonequilibrium free energy, which is defined as follows.

For a state $\rho\in \mathcal{D}(\mathcal{H}_S)$, the nonequilibrium free energy is defined by
\begin{align}
    F(\rho):=\Tr[\rho H_S]-\beta^{-1} S(\rho),
\end{align}
where $S(\rho)$ denotes the von Neumann entropy of $\rho$:
\begin{align}
    S(\rho):=-\Tr[\rho \log \rho] .
\end{align}
The equilibrium free energy  $F_\mathrm{eq}$ at inverse temperature $\beta$ is given by
\begin{align}
    F_\mathrm{eq}:=F(\gamma_S)=-\frac{1}{\beta}\log \Tr[e^{-\beta H_S}].
\end{align}
For two states $\rho, \sigma\in D(\mathcal{H}_S)$, the quantum relative entropy is defined by
\begin{align}\label{supp:eq:def-relative-entropy}
    D(\rho\|\sigma):= \Tr[\rho\log\rho-\rho\log\sigma].
\end{align}
Setting $\sigma = \gamma_S$ in Eq.~\eqref{supp:eq:def-relative-entropy}, we obtain
\begin{align}
    D(\rho\|\gamma_S) = \beta\left(F(\rho) - F_\mathrm{eq}\right).
\end{align}
Thus, the relative entropy with respect to the Gibbs state quantifies the excess free energy of $\rho$ above the equilibrium value. With a slight abuse of terminology, we henceforth refer to this quantity simply as the free energy.

The following proposition gives the necessary and sufficient condition for asymptotic conversions under Gibbs-preserving operations~\cite{supp:Buscemi-2019}.

\begin{proposition}[Characterization of asymptotic conversion under Gibbs-preserving operations]\label{supp:prop:GPO}
    Let $\rho, \sigma \in \mathcal{D}(\mathcal{H}_S)$. Then the optimal asymptotic conversion rate from $\rho$ to $\sigma$ under Gibbs-preserving operations is given by the free-energy ratio:
    \begin{align}
        R_\mathrm{GPO}(\rho\to\sigma) = \frac{D(\rho\|\gamma_S)}{D(\sigma\|\gamma_S)}.
    \end{align}
\end{proposition}

\begin{remark}
    If $\sigma=\gamma_S$, we adopt the convention that the free-energy ratio $\dfrac{D(\rho\|\gamma_S)}{D(\sigma\|\gamma_S)}$ is $\infty$, including when $\rho = \gamma_S$
\end{remark}

\subsubsection{Covariant operations}

Another important property of thermal operations is time-translation covariance, which we refer to simply as covariance. Namely, every thermal operation $\mathcal{E}:\mathcal{B}(\mathcal{H}_{S_1})\to\mathcal{B}(\mathcal{H}_{S_2})$ satisfies
\begin{align}\label{supp:eq:covariance-condition}
    \mathcal{E}\left(e^{-iH_{S_1}t}(\,\cdot\,) e^{iH_{S_1}t}\right)=
    e^{-iH_{S_2}t}\mathcal{E}(\,\cdot\,)e^{iH_{S_2}t}
\end{align}
for all $t\in\mathbb R$. Because of this property, thermal operations cannot break any time-translation symmetry possessed by the input state, which leads to restrictions on coherence manipulations.

Motivated by this property, we define covariant operations as follows. A CPTP map $\mathcal{E}:\mathcal{B}(\mathcal{H}_{S_1})\to\mathcal{B}(\mathcal{H}_{S_2})$ is a covariant operation if and only if $\mathcal{E}\left(e^{-iH_{S_1}t}(\,\cdot\,) e^{iH_{S_1}t}\right)=e^{-iH_{S_2}t}\mathcal{E}(\,\cdot\,)e^{iH_{S_2}t}$ for all $t\in\mathbb R$. By definition, every thermal operation is a covariant operation. This inclusion is strict: there exist covariant operations that are not thermal operations. For example, a CPTP map $\mathcal{E}(\,\cdot\,) = I_{S_2}/\mathrm{dim}S_2$ is covariant, but not Gibbs-preserving in general.

Asymptotic state conversion under covariant operations is defined analogously to that under thermal operations in Section~\ref{supp:subsubsec:asymptotic-TO}. The optimal conversion rate under covariant operations, denoted by $R_\mathrm{CO}(\rho\to\sigma)$, is defined as the supremum of all conversion rates achievable under covariant operations.

Covariant operations preserve every time-translation symmetry of the initial state. To formulate this property precisely, we introduce the symmetry subgroup as
\begin{align}
    \mathrm{Sym}(\rho):=\{t\in\mathbb{R}:\mathcal{U}_t (\rho) = \rho\},
\end{align}
where $\mathcal{U}_t(\,\cdot\,):=e^{-iH_S t}(\,\cdot\,)e^{iH_S t}$ is a time-evolution map generated by the Hamiltonian $H_S$ for time $t$. The covariance condition Eq.~\eqref{supp:eq:covariance-condition} implies that, under covariant operations, the symmetry subgroup of the initial state is always contained in that of the final state if the conversion is exact. Even in the asymptotic regime, where vanishing conversion error is allowed, no positive rate is achievable if the symmetry subgroup of the initial state is not contained in that of the target state. Namely, an asymptotic conversion from $\rho$ to $\sigma$ with a positive rate is possible only when $\mathrm{Sym}(\rho)\subseteq \mathrm{Sym}(\sigma)$~\cite{supp:Yamaguchi-2026-pure}. We henceforth call this requirement the symmetry condition.

For Hamiltonians whose energy-level spacings are integer multiples of a common constant, which we assume in this work, asymptotic conversions between pure states under covariant operations are characterized by the energy variance $V_{H_S}(\rho) := \Tr[H_S^2\rho] - \left(\Tr[H_S \rho]\right)^2$, together with the symmetry condition. The following proposition provides the necessary and sufficient condition for asymptotic conversions between pure states under covariant operations~\cite{supp:Marvian-2022,supp:Yamaguchi-2026-pure}.
 
\begin{proposition}[Characterization of asymptotic pure-state conversion under covariant operations]\label{supp:prop:CO}
    Suppose that all energy-level spacings of $H_S$ are integer multiples of some constant, and let $\psi, \phi \in \mathcal{D}(\mathcal{H}_S)$ be pure states.  Then the optimal asymptotic conversion rate from $\psi$ to $\phi$ under covariant operations is given by the energy-variance ratio when the symmetry condition holds, and is zero otherwise:
    \begin{align}\label{supp:eq:COrate}
        R_\mathrm{CO}(\psi\to\phi) =
        \begin{cases}
            \dfrac{V_{H_S}(\psi)}{V_{H_S}(\phi)} &  \text{if}\quad\mathrm{Sym}(\psi) \subseteq \mathrm{Sym}(\phi), \\
            0 & \text{otherwise}.
        \end{cases}
    \end{align}
\end{proposition}

\begin{remark}
    If $V_{H_S}(\phi)=0$, we adopt the convention that the variance ratio $\dfrac{V_{H_S}(\psi)}{V_{H_S}(\phi)}$ is $\infty$, including when $V_{H_S}(\psi)=0$.
\end{remark}
\begin{remark}
    Under the above assumption on the Hamiltonian, the time-translation symmetry is described by the group $U(1)$. In this case, the optimal conversion rate between pure states $R_\mathrm{CO}(\psi\to\phi)$ is given by Eq.\eqref{supp:eq:COrate}. For systems with general Hamiltonians, the optimal conversion rate $R_\mathrm{CO}(\psi\to\phi)$ is known to be characterized instead by the quantum geometric tensor, together with the symmetry condition \cite{supp:Yamaguchi-2026-pure}.
\end{remark}

\subsection{Asymptotic behavior of the energy distribution of i.i.d. copies of a pure state}

Energy distributions play a central role in our construction of thermal operations. Here we define the energy distribution of a pure state and explain its asymptotic behavior.

\subsubsection{Energy distribution of a pure state}

Let $\mathcal{H}_S$ be a $d$-dimensional Hilbert space and
\begin{align}
    H_S=\sum_{E} E\, \Pi_E
\end{align}
be the spectral decomposition of $H_S$, where $\Pi_E$ is a projector onto the energy eigenspace with energy $E$. The energy distribution of $\ket{\psi}\in\mathcal H_S$ is given by
\begin{align}
    p_\psi(E):=\mel{\psi}{\Pi_E}{\psi}.
\end{align}
We write the mean energy and energy variance of $\psi$ as
\begin{align}
    E_\psi := \mel{\psi}{H_S}{\psi} = \sum_E Ep_\psi(E)
\end{align}
and
\begin{align}
    V_{H_S}(\psi):=\mel{\psi}{H_S^2}{\psi}-\mel{\psi}{H_S}{\psi}^2 = \sum_E (E-E_\psi)^2p_\psi(E).
\end{align}

In the proof, we use the monotonicity of the total variation distance under convolution, as stated in the following lemma. The proof of Lemma~\ref{supp:lem:convolution-contraction} is given in Section~\ref{supp:sec:technical-lemmas}.

\begin{lemma}[Monotonicity of total variation distance under convolution]
\label{supp:lem:convolution-contraction}
For any probability distributions $p_1,p_2,q_1,q_2$ on $\mathbb Z$, we have
\begin{align}
    d_{\mathrm{TV}}\left(p_1*q_1,p_2*q_2\right)\leq
    d_{\mathrm{TV}}\left(p_1,p_2\right)
    +
    d_{\mathrm{TV}}\left(q_1,q_2\right).
    \label{supp:eq:convolution-tv-bound}
\end{align}
Here, for any probability distributions $p, q$, 
$d_\mathrm{TV}(p,q)$ denotes the total variation distance defined by
\begin{align}
    d_\mathrm{TV}(p,q):=\frac{1}{2}\sum_x|p(x)-q(x)|,
\end{align}
and
$p*q$ denotes the convolution of $p$ and $q$ defined by
\begin{align}
    (p*q)(x) := \sum_{y}p(y)q(x-y).
\end{align}

\end{lemma}

\subsubsection{Shifted Poisson distribution and translated Poisson distribution}

Energy distributions are important in constructing thermal operations because, when two pure states have close energy distributions, an energy-conserving unitary can map one state to a state close to the other~\cite{supp:Marvian-2013,supp:Marvian-2022}. This fact motivates us to describe the energy distribution of many copies of a pure state. To this end, we introduce two distributions called shifted Poisson distribution and translated Poisson distribution. 

We first define the shifted Poisson distribution. Let $s\in\mathbb Z$ and $v>0$. A random variable $X$ is said to follow the shifted Poisson distribution with shift parameter $s$ and Poisson parameter $v$, written as $X\sim SP(s,v)$, if
\begin{align}
    X=s+Y,
    \qquad
    Y\sim P(v),
\end{align}
where $P(v)$ is the Poisson distribution with parameter $v$. The mean and variance of $SP(s,v)$ are
\begin{align}
    \mathbb E[X]=s+v,
    \qquad 
    \operatorname{Var}(X)=v.
\end{align}
The sum of two independent random variables following shifted Poisson distributions also follows a shifted Poisson distribution. More precisely, if
\begin{align}
    X_1\sim SP(s_1,v_1),
    \qquad
    X_2\sim SP(s_2,v_2)
\end{align}
are independent, then
\begin{align}
    X_1+X_2\sim SP(s_1+s_2,v_1+v_2),
\end{align}
or equivalently,
\begin{align}
    SP(s_1,v_1)*SP(s_2,v_2)=SP(s_1+s_2,v_1+v_2).
    \label{supp:eq:sp-convolution-preliminaries}
\end{align}
In addition, two shifted Poisson distributions whose shift and Poisson parameters differ by $O(1)$ are close in total variation distance, as stated in the following lemma. The proof of Lemma~\ref{supp:lem:shifted-poisson} is given in Section~\ref{supp:sec:technical-lemmas}.

\begin{lemma}[Asymptotic convergence of shifted Poisson distributions]
\label{supp:lem:shifted-poisson}
Let $s_n,t_n\in\mathbb Z$ and $\lambda_n,\nu_n>0$. Suppose that
\begin{align}
    |s_n-t_n|=O(1),
    \qquad
    |\lambda_n-\nu_n|=O(1),
    \qquad
    \lambda_n,\nu_n=\Theta(n).
\label{supp:eq:sp-poisson-parameter-scaling}
\end{align}
Then the total variation distance between $SP(s_n,\lambda_n)$ and $SP(t_n,\nu_n)$ vanishes:
\begin{align}
    \lim_{n\to\infty}d_{\mathrm{TV}}\left(SP(s_n,\lambda_n),SP(t_n,\nu_n)\right)=0.
    \label{supp:eq:sp-asymptotic-equivalence}
\end{align}
\end{lemma}

We next define the translated Poisson distribution. Fix $\mu\in\mathbb R$ and $\sigma^2>0$, and define
\begin{align}
    s:=\left\lfloor\mu-\sigma^2\right\rfloor,\qquad\gamma:=\mu-\sigma^2-\left\lfloor\mu-\sigma^2\right\rfloor.
\end{align}
By definition, we have $0\le\gamma<1$. The translated Poisson distribution with parameters $\mu$ and $\sigma^2$, denoted by $TP(\mu,\sigma^2)$, is defined in terms of a shifted Poisson distribution by
\begin{align}
    TP(\mu,\sigma^2):=SP\left(s,\sigma^2+\gamma\right).
    \label{supp:eq:tp-as-sp-preliminaries}
\end{align}
Thus, if
\begin{align}
    Z\sim TP(\mu,\sigma^2),
\end{align}
then
\begin{align}
    Z=s+Y,\qquad 
    Y\sim P\left(\sigma^2+\gamma\right).
\end{align}
By construction, the mean of $Z$ is exactly $\mu$:
\begin{align}
    \mathbb E[Z]=s+\sigma^2+\gamma=\mu.
\end{align}
Its variance is
\begin{align}
    \operatorname{Var}(Z)=\sigma^2+\gamma,
\end{align}
and hence
\begin{align}
    \sigma^2\le\operatorname{Var}(Z)<\sigma^2+1.
\label{supp:eq:tp-variance-bound}
\end{align}
Therefore, $TP(\mu,\sigma^2)$ is an integer-valued distribution whose mean is exactly $\mu$, and whose variance differs from $\sigma^2$ by less than one.

\subsubsection{Energy distribution of i.i.d. copies of a pure state}

We review known results on the asymptotic behavior of the energy distribution for many copies of a pure state $\ket{\psi}$. Before considering the many-copy setting, we first describe the energy distribution of a single-copy pure state $\ket{\psi}\in\mathcal{H}_S$. We define the minimal period of a state as
\begin{align}
    \tau(\psi):= \mathrm{inf}\{t>0:\mathcal{U}_t(\psi)=\psi\}.
\end{align}
We assume that all energy-level spacings of $H_S$ are integer multiples of some constant. This assumption guarantees that every pure state $\psi$ on $S$ has a finite minimal period: $0\le\tau(\psi)<\infty$. If $\tau(\psi) = 0$, $\psi$ is an energy eigenstate, and hence the energy distributions of $\psi$ and $\psi^{\otimes n}$ are trivial. Thus, here we focus on the case where $\psi$ has a nontrivial minimal period: $\tau:=\tau(\psi)>0$.

The energy levels appearing in $\ket{\psi}$ lie on a lattice of spacing $E_\tau:=2\pi/\tau$. Let $E_\psi^{\min}$ be the minimum energy in the support of $\ket{\psi}$, that is,
\begin{align}\label{supp:eq:Emin-def}
    E_\psi^{\min}:=\min\{E:\mel{\psi}{\Pi_E}{\psi}>0\}.
\end{align}
Then, every energy level $E$ appearing in $\ket{\psi}$ can be written as
\begin{align}
    E=E_\psi^{\min}+kE_\tau,
    \qquad
    k\in\mathbb Z_{\ge 0}.
\end{align}
We define a probability distribution $\widetilde p_\psi$ on $\mathbb{Z}_{\ge 0}$ by
\begin{align}
    \widetilde p_\psi(k):=
    \mel{\psi}{\Pi_{E_\psi^{\min}+kE_\tau}}{\psi}.
\end{align}
If $E_\psi^{\min}+kE_\tau$ is not an eigenvalue of $H_S$, we set $\Pi_{E_\psi^{\min}+kE_\tau}=0$. Then the energy distribution of $\ket{\psi}$ is written as
\begin{align}
    p_\psi(E):=\mel{\psi}{\Pi_E}{\psi}=
    \begin{cases}
    \widetilde p_\psi(k)&\text{if}\quad E=E_\psi^{\min}+kE_\tau,\\
    0&\text{otherwise}.
    \end{cases}
\end{align}
Let $X_\psi$ be an integer-valued random variable with distribution $\widetilde p_\psi$. We write its mean and variance as
\begin{align}
    \mu_\psi:=\mathbb E[X_\psi],\qquad\sigma_\psi^2:=\mathrm{Var}(X_\psi).
\end{align}
These are related to the mean energy $E_\psi$ and the energy variance $V_{H_S}(\psi)$ by
\begin{align}
    \mu_\psi=\frac{E_\psi-E_\psi^{\min}}{E_\tau},\qquad
    \sigma_\psi^2=\frac{V_{H_S}(\psi)}{E_\tau^2}.
\end{align}
For $n$ copies, let $X_{\psi}^{(1)},\cdots,X_{\psi}^{(n)}$ be independent copies of $X_\psi$ and set
\begin{align}\label{supp:eq:sum-X}
    X_{\psi,n}:=X_{\psi}^{(1)}+\cdots+X_{\psi}^{(n)}.
\end{align}
Then the total energy of $\ket{\psi}^{\otimes n}$ is distributed as $nE_\psi^{\min}+E_\tau X_{\psi,n}$. Since $nE_\psi^\mathrm{min}$ and $E_\tau$ are deterministic, all nontrivial asymptotic behavior is contained in $X_{\psi,n}$. It therefore suffices to analyze the distribution of $X_{\psi,n}$. As stated in the following proposition, this distribution is asymptotically approximated by the translated Poisson distribution $TP(n\mu_\psi,n\sigma_\psi^2)$~\cite{supp:Marvian-2022,supp:Barbour-2002}.

\begin{proposition}[Translated Poisson approximation]\label{supp:prop:tp-approx}
Suppose that all energy-level spacings of $H_S$ are integer multiples of some constant, and let $\psi\in\mathcal{D}(\mathcal{H}_S)$ be a pure state with a positive minimal period $\tau(\psi)>0$. Then the distribution of $X_{\psi, n}$ defined in Eq.~\eqref{supp:eq:sum-X}, denoted by $\mathcal{L}(X_{\psi, n})$, converges to $TP( n\mu_\psi, n\sigma_\psi^2)$ in total variation distance:
\begin{align}
    \lim_{n\to\infty}d_{\mathrm{TV}}\left(\mathcal L(X_{\psi,n}),TP(n\mu_\psi,n\sigma_\psi^2)\right)=0.
\end{align}
\end{proposition}

Therefore, in the asymptotic limit, the energy distribution of $\psi^{\otimes n}$ is fully characterized by its mean and variance.


\section{Main result}\label{supp:sec:main-result}

We study the asymptotic conversion of pure states under thermal operations. To exclude the trivial case, we assume that $H_S$ has at least two distinct energy eigenvalues. Let
\begin{align}
    H_S=\sum_{i=1}^{d} E_i \ketbra{i}{i}
    \qquad
    (E_1\le E_2\le \cdots \le E_d)
\end{align}
be the spectral decomposition of $H_S$. In this work, we assume that all energy-level spacings of $H_S$ are integer multiples of some unit energy. By shifting the energy origin if necessary, we assume, without loss of generality, that the ground state energy is zero: $E_1=0$. Thus, there exist $E_*>0$ and $n_i\in\mathbb{Z}$ such that
\begin{align}
    E_i= E_* n_i,
    \qquad
    0=n_1\le n_2\le \cdots \le n_d.   
\end{align}
In addition, we can choose $E_*$ so that $n_1, \cdots ,n_d$ are coprime, i.e., $\gcd\{n_1,\cdots, n_d\} = 1$. (If $g = \gcd\{n_1, n_2,\cdots, n_d\}>1$, replacing $E_*$ by $gE_*$ and $n_i$ by $n_i/g$ yields $\gcd\{n_1,\cdots, n_d\} = 1$.)

In this setting, we derive the following necessary and sufficient condition for the asymptotic conversion of pure states under thermal operations, which is the main result of this work.

\begin{theorem}[Characterization of asymptotic pure-state conversion under thermal operations]
\label{supp:main}
Suppose that all energy-level spacings of $H_S$ are integer multiples of some constant, and let $\psi, \phi \in \mathcal{D}(\mathcal{H}_S)$ be pure states. Then the optimal asymptotic conversion rate from $\psi$ to $\phi$ is given by the minimum of the free-energy ratio and the energy-variance ratio when the symmetry condition holds, and is zero otherwise:
\begin{align}
    R_\mathrm{TO}(\psi\to\phi)
    =
    \begin{cases}
        \min\left\{
        \dfrac{D(\psi\|\gamma_S)}{D(\phi\|\gamma_S)},
        \dfrac{V_{H_S}(\psi)}{V_{H_S}(\phi)}
        \right\} &  \text{if}\quad\mathrm{Sym}(\psi) \subseteq \mathrm{Sym}(\phi), \\
        0 & \text{otherwise}.
    \end{cases}
\end{align}
\end{theorem}


\section{Proof of the converse part of the main theorem}\label{supp:sec:converse}

In this section, we give a proof of the converse part of Theorem~\ref{supp:main}, which follows directly from the fact that every thermal operation is both Gibbs-preserving and time-translation covariant.

Every conversion rate achievable under thermal operations is also achievable under Gibbs-preserving operations. Therefore, we have
\begin{align}\label{supp:eq:GPObound}
    R_\mathrm{TO}(\psi\to\phi)\le R_\mathrm{GPO}(\psi\to\phi).
\end{align}
Similarly, since every thermal operation is covariant, we have
\begin{align}\label{supp:eq:CObound}
    R_\mathrm{TO}(\psi\to\phi)\le R_\mathrm{CO}(\psi\to\phi).
\end{align}
Combining Eqs.~\eqref{supp:eq:GPObound} and \eqref{supp:eq:CObound}, we obtain
\begin{align}
    R_\mathrm{TO}(\psi\to\phi)\le\min\left\{R_\mathrm{GPO}(\psi\to\phi), R_\mathrm{CO}(\psi\to\phi)\right\}.
\end{align}
Substituting the expressions for $R_\mathrm{GPO}(\psi\to\phi)$ and $R_\mathrm{CO}(\psi\to\phi)$ given in Propositions \ref{supp:prop:GPO} and \ref{supp:prop:CO}, respectively, yields
\begin{align}
    R_\mathrm{TO}(\psi\to\phi)
    \le
    \begin{cases}
        \min\left\{
        \dfrac{D(\psi\|\gamma_S)}{D(\phi\|\gamma_S)},
        \dfrac{V_{H_S}(\psi)}{V_{H_S}(\phi)}
        \right\} &  \text{if}\quad\mathrm{Sym}(\psi) \subseteq \mathrm{Sym}(\phi), \\
        0 & \text{otherwise},
    \end{cases}
\end{align}
which proves the converse part of Theorem~\ref{supp:main}.

\section{Proof of the achievability part of the main theorem}\label{supp:sec:achievability}

In this section, we prove the achievability part of Theorem~\ref{supp:main} by constructing a sequence of thermal operations for any conversion rate $R$ below the claimed optimal rate. Our strategy is as follows. A positive conversion rate is possible only if the symmetry condition $\mathrm{Sym}(\psi)\subseteq \mathrm{Sym}(\phi)$ holds. It therefore suffices to consider pairs of states satisfying this condition. First, we consider the case $\mathrm{Sym}(\psi)= \mathrm{Sym}(\phi)$ and prove achievability by explicitly constructing a conversion protocol. We then consider the case $\mathrm{Sym}(\psi)\subsetneq \mathrm{Sym}(\phi)$ and reduce the problem to conversions between states with the same symmetry subgroup.

\subsection{Achievability when $\mathrm{Sym}(\psi)=\mathrm{Sym}(\phi)$}

\subsubsection{Overview of the protocol}

We give an overview of our construction of thermal operations for a pair of states with the same time-translation symmetry: $\mathrm{Sym}(\psi)= \mathrm{Sym}(\phi)$. In the construction, we mainly consider the case where $\psi$ and $\phi$ are not energy eigenstates and have nontrivial time-translation symmetry, i.e., $\mathrm{Sym}(\psi)= \mathrm{Sym}(\phi)\subsetneq \mathbb{R}$. For conversions between energy eigenstates, the constructed protocol still works with a few minor modifications, as we discuss later.

We construct a sequence of thermal operations for an arbitrary rate $0<R<R^*(\psi\to\phi)$ with
\begin{align}
    R^*(\psi\to\phi):=
    \min\left\{\frac{D(\psi\|\gamma_S)}{D(\phi\|\gamma_S)},\frac{V_{H_S}(\psi)}{V_{H_S}(\phi)}\right\}.
\end{align}
For each $n$, we set $m_n:=\lfloor Rn\rfloor$.

We first construct the auxiliary system. Motivated by the fact that two pure states can be approximately interconverted by an energy-conserving unitary if their energy distributions are close in total variation distance, we introduce the auxiliary system that adjusts the input and output energy distributions. The energy distributions of $\ket{\psi}^{\otimes n}$ and $\ket{\phi}^{\otimes m_n}$ are asymptotically characterized by their means and variances.  Here, $B$ serves to adjust the mean of the energy distribution and $B'$ serves to adjust its variance. The initial state of the total system is
\begin{align}
    \psi^{\otimes n}\otimes\gamma_B\otimes\gamma_{B'}.
\end{align}
The first auxiliary bath $B$ consists of many copies of the system $S$. Since its Gibbs state $\gamma_B$ is highly concentrated on typical energy eigenspaces, it suffices to construct the desired conversion for typical energy eigenstates of $B$. For each typical state $\ket{\lambda}$ of bath $B$, we choose a corresponding output energy eigenstate $\ket{t(\lambda)}\in\mathcal{H}_{\widetilde{B}}$. Note that $\mathcal{H}_S^{\otimes n}\otimes \mathcal{H}_B=\mathcal{H}_S^{\otimes m_n}\otimes \mathcal{H}_{\widetilde{B}}$ and therefore $\widetilde{B}$ is generally different from $B$. The second auxiliary bath $B'$ is a specially designed system such that the energy gaps between the ground state and excited states are very large and hence its Gibbs state is asymptotically concentrated on its ground state $\ket{g}$. We choose an output state $\ket{\eta_n}$ of $B'$ to be a superposition of excited energy eigenstates.

We first fix a typical energy eigenstate $\ket{\lambda}$ and show that, with appropriate choices of the bath states, the total variation distance between the energy distributions of $\ket{\Psi_\lambda}:=\ket{\psi}^{\otimes n}\ket{\lambda}\ket{g}$ and $\ket{\Phi_\lambda}:=\ket{\phi}^{\otimes m_n}\ket{t(\lambda)}\ket{\eta_n}$ vanishes asymptotically. The output state $\ket{t(\lambda)}\in\mathcal{H}_{\widetilde{B}}$ is constructed so that the difference between the mean energies of $\ket{\Psi_\lambda}$ and $\ket{\Phi_\lambda}$ is bounded by a constant. The state $\ket{\eta_n}\in\mathcal H_{B'}$ is constructed so that bath $B'$ absorbs the difference between the input and output energy variances. The latter variance adjustment is indeed possible because the condition 
\begin{align}
    R<R^*(\psi\to\phi)\le\frac{V_{H_S}(\psi)}{V_{H_S}(\phi)}
\end{align}
ensures that the additional variance in bath $B'$ is positive.

Next, we construct a single energy-conserving unitary $V$ that performs the desired conversion $\ket{\Psi_\lambda}\to\ket
{\Phi_\lambda}$ simultaneously for all typical bath states. We use the free-energy condition
\begin{align}
    R<R^*(\psi\to\phi)\le\frac{D(\psi\Vert\gamma_S)}{D(\phi\Vert\gamma_S)}
\end{align}
to show that there are sufficiently many states $\ket{t(\lambda)}$ with the properties required for the energy-distribution matching described above. This allows us to choose the target states $\{\ket{\Phi_\lambda}\}_{\lambda\in\mathcal{T}_\ell}$ to be mutually orthogonal, just as the input states $\{\ket{\Psi_\lambda}\}_{\lambda\in\mathcal{T}_\ell}$ are. Combining the above two facts, namely, the asymptotic vanishing of the total variation distance between the input and output energy distributions for each typical state and the orthogonality of the desired output states, we define a unitary $V_E$ in each energy sector. Then, we demonstrate that an energy-conserving unitary $V:= \bigoplus_EV_E$ performs the desired conversion with a vanishing error for all typical states. Finally, we show that the thermal operation defined by the two baths $B$ and $B'$ and the energy-conserving unitary $V$ achieves every conversion rate below $R^*(\psi\to\phi)$.

\subsubsection{Construction of the auxiliary system}\label{supp:subsec:baths}

In this subsection, we consider the case in which $\psi$ and $\phi$ have the same nontrivial time-translation symmetry: $\mathrm{Sym}(\psi) = \mathrm{Sym}(\phi)\subsetneq \mathbb{R}$. This condition implies that $\psi$ and $\phi$ have the same positive minimal period, i.e., $\tau:=\tau(\psi)=\tau(\phi)>0$. The case of $\mathrm{Sym}(\psi) = \mathrm{Sym}(\phi)=\mathbb{R}$ is considered later.

We first introduce two ancillary systems, denoted by $B$ and $B'$. The first bath $B$ consists of $\ell:=\lfloor n^{3/2}\rfloor$ copies of the original system $S$. Its Hamiltonian and Gibbs state are
\begin{align}
    H_B:=H_S^{(\ell)}=
    \sum_{j=1}^{\ell} I_S^{\otimes(j-1)}\otimes H_S\otimes I_S^{\otimes(\ell-j)},\qquad
    \gamma_B:=\gamma_S^{\otimes \ell}.
\end{align}
We expand the Gibbs state of $S$ as
\begin{align}
    \gamma_S=\sum_{i=1}^{d}g_i\ketbra{i}{i},\qquad
    g_i:=\frac{e^{-\beta E_i}}{\sum_{j=1}^{d}e^{-\beta E_j}},
\end{align}
and write the energy distribution of $\gamma_S$ as $\bm{g}:=(g_1,\ldots,g_d)$. An energy eigenstate of $B$ is written as
\begin{align}
    \ket{\bm{i}}:=\ket{i_1}\otimes\cdots\otimes\ket{i_\ell},\qquad
    i_j\in\{1,\ldots,d\}.
\end{align}
For $\bm i=(i_1,\ldots,i_\ell)$, let
\begin{align}
    N_a(\bm i):=\left|\left\{j\in\{1,\ldots,\ell\}:i_j=a\right\}\right|
\end{align}
be the occupation number of level $a$, and we define its occupation vector as
\begin{align}
    \bm{s}:=\left(N_1(\bm{i}),\ldots,N_d(\bm{i})\right).
\end{align}
We define the typical set $\mathcal{T}_\ell$ by
\begin{align}\label{supp:eq:typical-set}
    \mathcal{T}_\ell:=
    \left\{
        \bm{i}\in\{1,\ldots,d\}^{\ell}:
        \left| N_a(\bm{i})-\ell g_a\right|
        \le
        \sqrt{\ell}\log \ell\text{ for all }a\in \{1,\ldots, d\}
    \right\}.
\end{align}
Let
\begin{align}
    \Pi_{\mathcal T_\ell}
    :=
    \sum_{\bm i\in\mathcal T_\ell}
    \ketbra{\bm i}{\bm i}
\end{align}
be the projector onto the typical subspace, and define the weight of the typical subspace in $\gamma_B$ by
\begin{align}
    p_{\mathrm{typ}}:=
    \Tr\left[\Pi_{\mathcal T_\ell}\gamma_B\right].
\end{align}
Then, the probability $1-p_\mathrm{typ}$ of obtaining an atypical state vanishes in the asymptotic limit, which can be seen as follows. For each $a$, the random variable $N_a$ has mean $\ell g_a$ and variance $\ell g_a(1-g_a)$. Hence, by Chebyshev's inequality,
\begin{align}
    1-p_{\mathrm{typ}}\le
    \sum_{a=1}^{d}\Pr\left[|N_a-\ell g_a|>\sqrt{\ell}\log \ell\right]\le
    \sum_{a=1}^d\frac{g_a(1-g_a)}{(\log \ell)^2}\le
    \frac{1}{(\log \ell)^2},
\end{align}
which implies that the Gibbs state of $B$ is in the typical subspace with overwhelming probability:
\begin{align}
    \lim_{n\to\infty}p_{\mathrm{typ}}=1.
\end{align}
It is convenient to label a typical energy eigenstate of bath $B$ by
\begin{align}
    \lambda=(\bm s,\alpha),
\end{align}
where $\bm s$ is its occupation vector and $\alpha$ labels distinct basis states with the same occupation vector. We write the corresponding state as $\ket{\lambda}$.

We next explain the construction of the second bath $B'$. Define integers $\Delta_n$ and $N_n$ by
\begin{align}\label{supp:eq:deltaN-def}
    \Delta_n:=\left\lfloor n^{3/4}\right\rfloor,
    \qquad 
    N_n:=\left\lfloor\sqrt{n}\log n\right\rfloor,
\end{align}
which satisfy $\Delta_n \gg N_n$ for sufficiently large $n$. Using these integers, we define the Hamiltonian of $B'$ by
\begin{align}
    H_{B'} :=0\ketbra{g}{g}+E_\tau\sum_{k=-N_n}^{N_n}(\Delta_n+k)\ketbra{e;k}{e;k},
\end{align}
where $E_\tau := 2\pi/\tau$. Namely, bath $B'$ consists of a nondegenerate ground state $\ket{g}$ with zero energy and $2N_n+1$ excited states $\ket{e;k}$ whose excitation energies are around $E_\tau\Delta_n$. We define a set of integers corresponding to excited states of $B'$ as
\begin{align}\label{supp:eq:I_ndef}
    I_n := \{\Delta_n-N_n, \Delta_n-N_n+1, \ldots,\Delta_n+N_n\}.
\end{align}
Let $\gamma_{B'}$ be its Gibbs state and let
\begin{align}\label{supp:eq:pgr-def}
    p_{\mathrm{gr}}:=\mel{g}{\gamma_{B'}}{g}
\end{align}
be the Gibbs weight of the ground state. Then, the  total Gibbs weight of the excited states $1-p_{\mathrm{gr}}$ satisfies
\begin{align}
    1-p_{\mathrm{gr}} \le (2N_n+1)\exp\left[-\beta E_\tau(\Delta_n-N_n)\right],
\end{align}
and therefore
\begin{align}
    \lim_{n\to\infty}p_{\mathrm{gr}}=1.
\end{align}
Hence, the Gibbs state of $B'$ converges in trace distance to the ground state $\ketbra{g}{g}$.

\subsubsection{Construction of the output states of bath $B$ and $B'$}

In our construction of a thermal operation, we need an energy-conserving unitary that performs the desired conversion for all typical bath states. Based on the fact that one pure state can be converted to another by an energy-conserving unitary if their energy distributions are close in total variation distance, we first fix a typical label $\lambda\in\mathcal{T}_\ell$, and, by introducing two output states $\ket{t(\lambda)}\in\mathcal{H}_{\widetilde{B}}$ and $\ket{\eta_n}\in\mathcal{H}_{B'}$, we compare the energy distributions of input and output pure states
\begin{align}
    \ket{\Psi_\lambda}:=\ket{\psi}^{\otimes n}\ket{\lambda}\ket{g}
    \in 
    \mathcal{H}_S^{\otimes n}\otimes \mathcal{H}_B \otimes \mathcal{H}_{B'},
    \qquad
    \ket{\Phi_\lambda}:=\ket{\phi}^{\otimes m_n}\ket{t(\lambda)}\ket{\eta_n}
    \in
    \mathcal{H}_S^{\otimes m_n} \otimes \mathcal{H}_{\widetilde{B}}\otimes \mathcal{H}_{B'}.
\end{align}
Here, $\widetilde{B}$ consists of $\ell':=n+\ell-m_n$ copies of $S$, and $\ket{t(\lambda)}\in\mathcal{H}_{\widetilde{B}}$ is an energy eigenstate of the output bath $\widetilde{B}$ and $\ket{\eta_n}\in\mathcal{H}_{B'}$ is an output state of bath $B'$. 

In this subsection, we construct the output states $\ket{t(\lambda)}$ and $\ket{\eta_n}$ with properties needed to show that the total variation distance between the energy distributions of $\ket{\Psi_\lambda}$ and $\ket{\Phi_\lambda}$ vanishes. The two baths play different roles. The state $\ket{t(\lambda)}$ compensates the difference between the input and output mean energies, and the state $\ket{\eta_n}$, which is independent of $\lambda$, absorbs the excess energy variance of the input.

\begin{center}
    \textit{Construction of the output state of bath \texorpdfstring{$B'$}{B-prime}}
\end{center}

We first construct the output state $\ket{\eta_n}$ of bath $B'$. The role of this bath is to absorb the excess energy variance in the system. We define the difference between the input and output energy variances, normalized by $E_\tau^2$, as
\begin{align}
    \Sigma_n^2:=\frac{nV_{H_S}(\psi)-m_nV_{H_S}(\phi)}{E_\tau^2}.
\end{align}
Since we have fixed a conversion rate $R$ satisfying $R < \dfrac{V_{H_S}(\psi)}{V_{H_S}(\phi)}$, we have $\Sigma_n^2>0$.

We seek to construct a state $\ket{\eta_n}$ whose energy distribution approximates a shifted Poisson distribution with variance approximately $E_\tau^2\Sigma_n^2$. For each finite $n$, however, the Hamiltonian of $B'$ has a bounded spectrum, whereas a shifted Poisson distribution has semi-infinite support. A shifted Poisson distribution therefore cannot be realized exactly as the energy distribution of a state on $B'$. We thus introduce a shifted Poisson distribution $q_n$ and construct $\ket{\eta_n}$ by truncating $q_n$ to the spectral range of $B'$ and renormalizing it. We then show that the truncated tail has vanishing probability as $n\to\infty$.

Let $\widetilde{q}_n$ be the translated Poisson distribution $TP(\Delta_n, \Sigma_n^2)$. Using $\widetilde{q}_n$, we introduce $q_n$ as an energy distribution corresponding to $\widetilde{q}_n$:
\begin{align}
    q_n(E) :=
    \begin{cases}
        \widetilde{q}_n(k) &\text{if}\quad  E = E_\tau k,\\
        0 & \text{otherwise}.
    \end{cases}
\end{align}
Then, $q_n$ is a shifted Poisson distribution with mean $E_\tau \Delta_n$ and variance approximately $E_\tau^2\Sigma_n^2$. The following lemma ensures the existence of a state $\ket{\eta_n}$ whose energy distribution approximates $q_n$.

\begin{lemma}[Construction of the output state of bath $B'$]
\label{supp:lem:eta-construction}
There exists a pure state $\ket{\eta_n}\in\mathcal H_{B'}$ such that
\begin{align}
    \lim_{n\to\infty}d_{\mathrm{TV}}\left(p_{\eta_n},q_n\right)=0,
\end{align}
where $p_{\eta_n}$ denotes the energy distribution of
$\ket{\eta_n}$.
\end{lemma}

\begin{proof}
We construct $\ket{\eta_n}$ as a superposition of excited energy eigenstates. First, we show that the probability weight of $\widetilde{q}_n$ outside $I_n$ (the set of integers corresponding to the spectrum of $H_{B'}$, introduced in Eq.~\eqref{supp:eq:I_ndef}) is negligible. We then restrict $\widetilde{q}_n$ to $I_n$ and renormalize the resulting distribution. 

Since the variance of $\widetilde{q}_n$ differs from $\Sigma_n^2$ by less than one, Chebyshev's inequality implies that the probability that a random variable drawn from $\widetilde{q}_n$ is not in $I_n$ is negligible:
\begin{align}
    \epsilon_n^{\mathrm{cut}}:=
    \sum_{z\notin I_n} \widetilde{q}_n(z)\le
    \frac{\Sigma_n^2+1}{N_n^2}=
    O\left(\frac{1}{(\log n)^2}\right).
\end{align}
We then introduce a truncated and normalized distribution supported on $I_n$ as
\begin{align}
    \widetilde p_{\eta_n}(z):=
    \begin{cases}
        \dfrac{\widetilde{q}_n(z)}{1-\epsilon_n^{\mathrm{cut}}}&\text{if}\quad z\in I_n,\\
        0&\text{otherwise}.
    \end{cases}
\end{align}
A direct calculation gives
\begin{align}
    d_{\mathrm{TV}}\left(\widetilde p_{\eta_n},\widetilde{q}_n\right)=\epsilon_n^{\mathrm{cut}}.
\end{align}
Using this $\widetilde{p}_{\eta_n}$, we define the final pure state of $B'$ as
\begin{align}
    \ket{\eta_n}:=\sum_{k=-N_n}^{N_n}\sqrt{\widetilde p_{\eta_n}(\Delta_n+k)}\ket{e;k}.
\end{align}
Let $p_{\eta_n}$ be the energy distribution of $\ket{\eta_n}$. Since $d_{\mathrm{TV}}\left(p_{\eta_n},q_n\right) = d_{\mathrm{TV}}\left(\widetilde{p}_{\eta_n},\widetilde{q}_n\right)$, we get
\begin{align}
    \lim_{n\to\infty}d_{\mathrm{TV}}\left(p_{\eta_n},q_n\right)
    =\lim_{n\to\infty}\epsilon_n^{\mathrm{cut}}=0,
\end{align}
which completes the proof.
\end{proof}

We henceforth take $\ket{\eta_n}$ to be the state constructed in Lemma~\ref{supp:lem:eta-construction}. Then, the state $\ket{\eta_n}$ absorbs the excess energy variance of the input state relative to the output state.

\begin{center}
    \textit{Construction of the output state of bath \texorpdfstring{$\widetilde{B}$}{B}}
\end{center}

We next construct the output state of bath $\widetilde{B}$ for each typical bath state $\lambda=(\bm s,\alpha)\in\mathcal T_\ell$. Since our input system consists of $n+\ell$ copies of $S$ and the desired output system consists of $m_n$ copies, the output bath $\widetilde{B}$ consists of $\ell':=n+\ell-m_n$ copies of $S$. Let
\begin{align}
    \bm{E}:=(E_1,\ldots,E_d)
\end{align}
be the vector of energy eigenvalues of $H_S$. Denoting an output occupation vector by
\begin{align}
    \bm{t}=(t_1,\ldots,t_d),
\end{align}
we define the input and output energy-lattice offsets by
\begin{align}
    A_{\lambda}^{\mathrm{in}}:=nE_{\psi}^{\min}+\bm{E}\cdot\bm{s},\qquad
    A_{\lambda}^{\mathrm{out}}:=m_nE_{\phi}^{\min}+\bm{E}\cdot\bm{t},
\end{align}
where $E_\psi^\mathrm{min}$ and $E_\phi^\mathrm{min}$ are the minimum energies in the supports of $\psi$ and $\phi$, respectively, as defined in Eq.~\eqref{supp:eq:Emin-def}. These quantities specify the offsets of the energy lattices associated with $\ket{\Psi_\lambda}$ and $\ket{\Phi_\lambda}$, so that their energy distributions are supported on $A_{\lambda}^{\mathrm{in}} + E_\tau \mathbb Z$ and $A_{\lambda}^{\mathrm{out}} + E_\tau \mathbb Z$, respectively. 

We require the occupation vector $\bm{t}$ to satisfy two conditions. 

The first condition, which we refer to as the energy-lattice condition, requires the energy-lattice offsets of the input and output to differ by an integer multiple of $E_\tau$. This condition ensures that the energy distributions of the input and output states are supported on the same lattice with spacing $E_\tau$. To illustrate why the energy-lattice condition is necessary, consider a three-level system with energy eigenstates $\ket{0}$, $\ket{1}$, and $\ket{2}$, whose energies are $0$, $E_*$, and $2E_*$, respectively. Suppose that $\ket{\psi}$ and $\ket{\phi}$ are superpositions of $\ket{0}$ and $\ket{2}$. Their energy distributions are then supported on the lattice $2E_*\mathbb{Z}$. By contrast, the Gibbs state $\gamma_S$ has full support on $\mathcal{H}_S$, so the energies of the typical energy eigenstates of $\gamma_S^{\otimes\ell}$ generally lie in $E_*\mathbb{Z}$. Consequently, it may happen that the energy distribution of $\ket{\Psi_\lambda}$ is supported on $\{0,2E_*,4E_*,\ldots\}$, whereas that of $\ket{\Phi_\lambda}$ is supported on $\{E_*,3E_*,5E_*,\ldots\}$. Since these two sets are disjoint, no energy-conserving unitary can convert $\ket{\Psi_\lambda}$ into $\ket{\Phi_\lambda}$. The energy-lattice condition rules out such a mismatch by ensuring that the two energy distributions are supported on the same lattice. We note that this condition is automatically satisfied whenever $E_\tau = E_*$, and in particular, this is always the case for states in a two-level system satisfying $\mathrm{Sym}(\psi) = \mathrm{Sym}(\phi)\subsetneq \mathbb{R}$.

The second requirement is that the mean energies of $\ket{\Psi_\lambda}$ and $\ket{\Phi_\lambda}$ differ by at most $E_\tau/2$, up to the truncation error arising from the construction of $\ket{\eta_n}$. We refer to this condition as the mean-energy condition. The following lemma ensures the existence of an occupation vector $\bm{t}$ satisfying both conditions.

\begin{lemma}[Construction of the output occupation vectors]
\label{supp:lem:output-occupation-vector}
For sufficiently large $n$, there exists $\bm{c}\in\mathbb{Z}^d$ such that, for every typical occupation vector $\bm{s}$ of bath $B$, $\bm{t}=\bm{s}+\bm{c}$ is a valid output occupation vector of $\widetilde{B}$ satisfying the following conditions.
\begin{enumerate}
    \item \text{(Energy-lattice condition)}
    The energy offsets of the input and output states satisfy
    \begin{align}\label{supp:eq:condition-2}
        A_{\lambda}^{\mathrm{out}}-A_{\lambda}^{\mathrm{in}}
        \in
        E_\tau\mathbb{Z}.
    \end{align}

    \item \text{(Mean-energy condition)}
    The mean energies of the input and output states satisfy
    \begin{align}\label{supp:eq:condition-3} 
        \left|nE_\psi - (m_nE_\phi + \bm{E}\cdot\bm{c} + E_\tau\Delta_n ) \right| 
        \le \frac{E_\tau}{2}.
    \end{align}
\end{enumerate}
\end{lemma}

\begin{proof}  
    Suppose that $a_n \in\mathbb Z$ and $\bm b^{(n)} \in \mathbb{Z}^d$ satisfy the following conditions.
    \begin{align}
        &\left|n\mu_\psi-m_n\mu_\phi-a_n-\Delta_n\right|\le\frac{1}{2}
        \label{supp:eq:an-mean-condition}\\
        &\sum_{i=1}^d b_i^{(n)}=n-m_n
        \label{supp:eq:bn-number-condition}\\
        &\bm E\cdot\bm b^{(n)}=nE_\psi^{\min}-m_nE_\phi^{\min}+a_nE_\tau
        \label{supp:eq:bn-energy-condition}
    \end{align}
    Here, $\mu_\psi$ and $\mu_\phi$ are defined through
    \begin{align}
        E_\psi&=E_\psi^{\min}+E_\tau\mu_\psi,\qquad
        E_\phi=E_\phi^{\min}+E_\tau\mu_\phi.
    \end{align}
    Before showing the existence of such $a_n$ and $\bm b^{(n)}$, we show that $\bm{c} = \bm{b}^{(n)}$ satisfies the two conditions~\eqref{supp:eq:condition-2} and~\eqref{supp:eq:condition-3}. We first verify the energy-lattice condition~\eqref{supp:eq:condition-2}. Using Eq.~\eqref{supp:eq:bn-energy-condition}, we obtain
    \begin{align}
        A_{\lambda}^{\mathrm{out}}-A_{\lambda}^{\mathrm{in}}
        &=
        m_nE_\phi^{\min}+\bm E\cdot\bm t-nE_\psi^{\min}-\bm E\cdot\bm s\\
        &=
        m_nE_\phi^{\min}-nE_\psi^{\min}+\bm E\cdot\bm b^{(n)}\\
        &=a_nE_\tau.\label{supp:eq:lattice-offset-from-an}
    \end{align}
    Since $a_n\in\mathbb Z$, this proves the energy-lattice condition~\eqref{supp:eq:condition-2}. Next, we verify the mean-energy condition~\eqref{supp:eq:condition-3}. Using Eq.~\eqref{supp:eq:bn-energy-condition},
    \begin{align}
        nE_\psi+\bm E\cdot\bm s-\left(m_nE_\phi+\bm E\cdot\bm t+E_\tau\Delta_n\right)
        &=
        nE_\psi-m_nE_\phi-\bm E\cdot\bm b^{(n)}-E_\tau\Delta_n\\
        &=
        n\left(E_\psi-E_\psi^{\min}\right)-m_n\left(E_\phi-E_\phi^{\min}\right)-E_\tau\left(a_n+\Delta_n\right)\\
        &=
        E_\tau\left(n\mu_\psi-m_n\mu_\phi-a_n-\Delta_n\right).
        \label{supp:eq:mean-difference-from-an}
    \end{align}
    It therefore follows from Eq.~\eqref{supp:eq:an-mean-condition} that
    \begin{align}
        \left|nE_\psi+\bm E\cdot\bm s-\left(m_nE_\phi+\bm E\cdot\bm t+E_\tau\Delta_n\right)\right|\le
        \frac{E_\tau}{2}.
    \end{align}
    Thus, the mean-energy condition~\eqref{supp:eq:condition-3} is also satisfied.
    
    We next show that $a_n$ and $\bm b^{(n)}$ satisfying Eqs.~\eqref{supp:eq:an-mean-condition}--\eqref{supp:eq:bn-energy-condition} exist. We first set
    \begin{align}
        a_n:=\left \lfloor n\mu_\psi-m_n\mu_\phi-\Delta_n+\frac{1}{2}\right\rfloor,\label{supp:eq:definition-an}
    \end{align}
    which fulfills Eq.~\eqref{supp:eq:an-mean-condition}. We next construct $\bm{b}^{(n)}$. Recall that the energy levels can be written as integer multiples of a unit energy $E_*$:
    \begin{align}
        E_i=E_*n_i,
    \end{align}
    where $n_i\in\mathbb Z$, $n_1=0$, and $\gcd\left(n_1,\cdots,n_d\right)=1$. We write
    \begin{align}
        E_\psi^{\min}=E_*e_\psi,
        \qquad
        E_\phi^{\min}=E_*e_\phi,
        \qquad
        E_\tau = E_* \kappa 
    \end{align}
    where $e_\psi,e_\phi\in\mathbb{Z}_{\ge 0}$ and $\kappa\in\mathbb Z_{>0}$. Eq.~\eqref{supp:eq:bn-energy-condition} is then equivalent to
    \begin{align}
        \sum_{i=1}^d n_i b_i^{(n)}=L_n,\qquad
        L_n:=ne_\psi-m_ne_\phi+\kappa a_n.
    \label{supp:eq:bn-weighted-sum}
    \end{align}
    Since $\gcd(n_1,\ldots,n_d)=1$,  there exist integers $z_1,\cdots,z_d$ such that
    \begin{align}
        \sum_{i=1}^d n_iz_i=1.
    \end{align}
    Now we set
    \begin{align}
        b_1^{(n)}&:=n-m_n-L_n\sum_{j=2}^d z_j,\\
        b_i^{(n)}&:=L_n z_i,\qquad i=2,\cdots,d.
    \end{align}
    It then follows that
    \begin{align}
        \sum_{i=1}^d b_i^{(n)}=n-m_n
    \end{align}
    and
    \begin{align}
        \sum_{i=1}^d n_ib_i^{(n)}=L_n,
    \end{align}
    where we used $n_1 = 0$. Thus Eqs.~\eqref{supp:eq:bn-number-condition} and~\eqref{supp:eq:bn-energy-condition} are satisfied. 
    
    Finally, we verify that $\bm t:=\bm s + \bm b^{(n)}$ is a valid occupation vector of $\widetilde{B}=S^{\otimes (n+\ell-m_n)}$. The choice of $a_n$ implies $a_n=O(n)$, and consequently $L_n=O(n)$. It follows from the above construction that $b_i^{(n)}=O(n)$ for every $i$. On the other hand, for every typical occupation vector $\bm s$, each component $s_i$ is of order $\ell = \lfloor n^{3/2}\rfloor$, since all Gibbs probabilities are strictly positive. Thus
    \begin{align}
        t_i=s_i+b_i^{(n)}\ge 0
        \end{align}
    holds for every $i$ and all sufficiently large $n$. In addition, Eq.~\eqref{supp:eq:bn-number-condition} implies
    \begin{align}
        \sum_{i=1}^d t_i = \sum_{i=1}^d (s_i + b_i^{(n)}) = n+\ell-m_n = \ell'.
    \end{align}
    Thus, $\bm t$ is a valid occupation vector of $\widetilde B$. This completes the proof.
\end{proof}

Below, for each typical bath state $\lambda=(\bm{s},\alpha)$, we denote by $\ket{t(\lambda)}$ an energy eigenstate of the output bath $\widetilde B$ with label $t(\lambda)=(\bm{t},\alpha')$, where $\bm{t}$ is the occupation vector constructed in Lemma~\ref{supp:lem:output-occupation-vector} and $\alpha'$ is a degeneracy label. We leave $\alpha'$ unspecified for the moment, and choose it later in Proposition~\ref{supp:prop:orthogonal-output-assignment} so that the output states $\{\ket{t(\lambda)}\}_{\lambda\in T_\ell}$ are mutually orthogonal.

\subsubsection{Comparison of the total energy distributions}

We have constructed the output states $\ket{\eta_n}$ and $\ket{t(\lambda)}$ for each typical bath state $\lambda\in\mathcal T_\ell$. We now show that the total energy distributions of $\ket{\Psi_\lambda}:=\ket{\psi}^{\otimes n}\ket{\lambda}\ket{g}$ and $\ket{\Phi_\lambda}:=\ket{\phi}^{\otimes m_n}\ket{t(\lambda)}\ket{\eta_n}$ are asymptotically close in total variation distance, uniformly over $\lambda\in\mathcal T_\ell$. The proof consists of two steps. First, by representing both energy distributions on a common integer lattice, we approximate them by two shifted Poisson distributions, denoted by $Q_n^{\mathrm{in}}$ and $Q_n^{\mathrm{out}}$. Then, we show that the distance between $Q_n^{\mathrm{in}}$ and $Q_n^{\mathrm{out}}$ vanishes asymptotically.

Defining dimensionless variances of $\psi$ and $\phi$ by
\begin{align}
    \sigma_\psi^2:=\frac{V_{H_S}(\psi)}{E_\tau^2},
    \qquad
    \sigma_\phi^2:=\frac{V_{H_S}(\phi)}{E_\tau^2},
    \end{align}
we have
    \begin{align}
    \Sigma_n^2
    =
    n\sigma_\psi^2-m_n\sigma_\phi^2.
    \label{supp:eq:sigma-difference-dimensionless}
\end{align}
By Eq.\eqref{supp:eq:condition-2}, the energy distributions of $\ket{\Psi_\lambda} = \ket{\psi}^{\otimes n}\ket{\lambda}\ket{g}$ and $\ket{\Phi_\lambda}=\ket{\phi}^{\otimes m_n} \ket{t(\lambda)}\ket{\eta_n}$ are supported on the common lattice $A_\lambda^{\mathrm{in}}+E_\tau\mathbb{Z}$. We define distributions on integers corresponding to energy distributions of $\ket{{\Psi}_\lambda}$ and $\ket{{\Phi}_\lambda}$ as
\begin{align}
    \widetilde P_{\Psi_\lambda}(k)
    &:=
    P_{\Psi_\lambda}\left(A_\lambda^{\mathrm{in}}+E_\tau k\right),\qquad
    \widetilde P_{\Phi_\lambda}(k)
    :=
    P_{\Phi_\lambda}\left(A_\lambda^{\mathrm{in}}+E_\tau k\right).
    \label{supp:eq:integer-total-energy-distributions}
\end{align}
Let $Y_n$ be an integer-valued random variable whose distribution is given by
\begin{align}
    \Pr[Y_n=k]:=p_{\eta_n}(E_\tau k).
    \label{supp:eq:Yn-definition}
\end{align}
Recalling that $a_n$, introduced in Eq.\eqref{supp:eq:definition-an}, satisfies $A_\lambda^\mathrm{out} - A_\lambda^\mathrm{in} = a_nE_\tau$, we have
\begin{align}
    \widetilde{P}_{\Psi_\lambda}=\mathcal{L}(X_{\psi,n}),\qquad
    \widetilde{P}_{\Phi_\lambda}=\mathcal{L}\left(a_n+X_{\phi,m_n}+Y_n\right),
    \label{supp:eq:integer-distribution}
\end{align}
where $\mathcal{L}(Z)$ denotes the distribution of a random variable $Z$, and $X_{\psi, n}$ and $X_{\phi, m_n}$ are random variables defined in Eq.~\eqref{supp:eq:sum-X}. Let $\delta_{a_n}$ be an integer-valued probability distribution concentrated at $a_n$. We define two distributions $Q_n^{\mathrm{in}}$ and $Q_n^{\mathrm{out}}$ on integers  by
\begin{align}
    Q_n^{\mathrm{in}}
    :=
    TP\left(n\mu_\psi,n\sigma_\psi^2\right),\qquad
    Q_n^{\mathrm{out}}
    :=
    \delta_{a_n}*TP\left(m_n\mu_\phi,m_n\sigma_\phi^2\right)*TP\left(\Delta_n,\Sigma_n^2\right).
    \label{supp:eq:ideal-tp}
\end{align}
Now we show that the energy distributions of the input and output states $\ket{\Psi_\lambda}$ and $\ket{\Phi_\lambda}$ converge to $Q_n^{\mathrm{in}}$ and $Q_n^{\mathrm{out}}$, respectively, as stated in the following lemma.

\begin{lemma}[Translated Poisson approximations of the total energy distributions]
\label{supp:lem:input-output-tp-approximation}
The integer-valued energy distributions of the input and output states satisfy
\begin{align}
    \lim_{n\to\infty}\sup_{\lambda\in\mathcal T_\ell}
    d_{\mathrm{TV}}\left(\widetilde P_{\Psi_\lambda},Q_n^{\mathrm{in}}\right)
    =0
    \label{supp:eq:input-tp-approximation}
\end{align}
and
\begin{align}
    \lim_{n\to\infty}\sup_{\lambda\in\mathcal T_\ell}
    d_{\mathrm{TV}}\left(\widetilde P_{\Phi_\lambda},Q_n^{\mathrm{out}}\right)
    =0.
    \label{supp:eq:output-tp-approximation}
\end{align}
\end{lemma}

\begin{proof}
We first consider the input state $\ket{\Psi_\lambda}$. By Eq.~\eqref{supp:eq:integer-distribution} and Proposition~\ref{supp:prop:tp-approx}, we have
\begin{align}
    \lim_{n\to\infty}d_{\mathrm{TV}}\left(\widetilde P_{\Psi_\lambda},Q_n^{\mathrm{in}}\right)
    =\lim_{n\to\infty}d_{\mathrm{TV}}\left(\mathcal L(X_{\psi,n}),TP\left(n\mu_\psi,n\sigma_\psi^2\right)\right)
    =0,
\end{align}
where $d_{\mathrm{TV}}(\mathcal L(X_{\psi,n}),TP(n\mu_\psi,n\sigma_\psi^2))$ is independent of $\lambda\in\mathcal{T}_\ell$. This completes the proof of Eq.~\eqref{supp:eq:input-tp-approximation}. We next consider the output state $\ket{\Phi_\lambda}$. Proposition~\ref{supp:prop:tp-approx} and Lemma~\ref{supp:lem:eta-construction} give
\begin{align}
    \lim_{n\to\infty}d_{\mathrm{TV}}\left(\mathcal L(X_{\phi,m_n}),TP\left(m_n\mu_\phi,m_n\sigma_\phi^2\right)\right)
    =0.
    \label{supp:eq:phi-tp-approximation}
\end{align}
and
\begin{align}
    \lim_{n\to\infty}d_{\mathrm{TV}}\left(\mathcal L(Y_n),TP\left(\Delta_n,\Sigma_n^2\right)\right)
    =0.
    \label{supp:eq:eta-tp-approximation-integer}
\end{align}
Since Eq.~\eqref{supp:eq:integer-distribution} reads
\begin{align}
    \widetilde P_{\Phi_\lambda}=\delta_{a_n}*\mathcal L(X_{\phi,m_n})* \mathcal L(Y_n),
    \label{supp:eq:output-distribution-convolution}
\end{align}
using the monotonicity of total variation distance under convolution (Lemma~\ref{supp:lem:convolution-contraction}), we have
\begin{align}\label{supp:eq:output-bound}
    d_{\mathrm{TV}}\left(\widetilde P_{\Phi_\lambda},Q_n^{\mathrm{out}}\right)
    &\le
    d_{\mathrm{TV}}\left(\mathcal L(X_{\phi,m_n}),TP\left(m_n\mu_\phi,m_n\sigma_\phi^2\right)\right)
    +
    d_{\mathrm{TV}}\left(\mathcal L(Y_n),TP\left(\Delta_n,\Sigma_n^2\right)\right),
\end{align}
where the upper bound is independent of $\lambda\in\mathcal{T}_\ell$. Combining Eqs.~\eqref{supp:eq:phi-tp-approximation},~\eqref{supp:eq:eta-tp-approximation-integer}, and~\eqref{supp:eq:output-bound}, we obtain Eq.~\eqref{supp:eq:output-tp-approximation}.
\end{proof}

We next argue that the total variation distance between $Q_n^\mathrm{in}$ and $Q_n^\mathrm{out}$ vanishes as $n\to\infty$. The proof of Lemma~\ref{supp:lem:input-output-ideal} is given in Section~\ref{supp:sec:technical-lemmas}.

\begin{lemma}[Convergence of $Q_n^\mathrm{in}$ and $Q_n^\mathrm{out}$]
\label{supp:lem:input-output-ideal}
The input and output distributions
$Q_n^{\mathrm{in}}$ and $Q_n^{\mathrm{out}}$ satisfy
\begin{align}
    \lim_{n\to\infty}d_{\mathrm{TV}}\left(Q_n^{\mathrm{in}},Q_n^{\mathrm{out}}\right)
    =0.
\end{align}
\end{lemma}

We now combine the two preceding lemmas to show the convergence of the input and output energy distributions in total variation distance.

\begin{proposition}[Uniform convergence of the input and output energy distributions]
\label{supp:prop:input-output-energy-distributions}
The total energy distributions of $\ket{\Psi_\lambda}$ and
$\ket{\Phi_\lambda}$ satisfy
\begin{align}
    \lim_{n\to\infty}\sup_{\lambda\in\mathcal T_\ell}d_{\mathrm{TV}}\left(P_{\Psi_\lambda},P_{\Phi_\lambda}\right)=0.
    \label{supp:eq:input-output-energy-distribution-convergence}
\end{align}
\end{proposition}

\begin{proof}
The triangle inequality gives
\begin{align}
    d_{\mathrm{TV}}\left(\widetilde P_{\Psi_\lambda},\widetilde P_{\Phi_\lambda}\right)
    \le
    d_{\mathrm{TV}}\left(\widetilde P_{\Psi_\lambda},Q_n^{\mathrm{in}}\right)
    +
    d_{\mathrm{TV}}\left(Q_n^{\mathrm{in}},Q_n^{\mathrm{out}}\right)
    +
    d_{\mathrm{TV}}\left(Q_n^{\mathrm{out}},\widetilde P_{\Phi_\lambda}\right).
    \label{supp:eq:final-energy-distribution-triangle}
\end{align}
By Lemma~\ref{supp:lem:input-output-tp-approximation} and Lemma~\ref{supp:lem:input-output-ideal}, each term on the right-hand side converges to zero uniformly over $\lambda\in\mathcal T_\ell$. Hence, we have
\begin{align}
    \lim_{n\to\infty}\sup_{\lambda\in\mathcal T_\ell}d_{\mathrm{TV}}\left(\widetilde P_{\Psi_\lambda},\widetilde P_{\Phi_\lambda}\right)
    =0.
\end{align}
Since $\widetilde{P}_{\Psi_\lambda}$ and $\widetilde{P}_{\Phi_\lambda}$ represent energy distributions on the same lattice $A_\lambda^{\mathrm{in}}+E_\tau\mathbb Z$, their total variation distance is exactly the same as the total variation distance between $P_{\Psi_\lambda}$ and $P_{\Phi_\lambda}$. Therefore, we obtain
\begin{align}
    \lim_{n\to\infty}\sup_{\lambda\in\mathcal T_\ell}d_{\mathrm{TV}}\left(P_{\Psi_\lambda},P_{\Phi_\lambda}\right)
    =
    \lim_{n\to\infty}\sup_{\lambda\in\mathcal T_\ell}d_{\mathrm{TV}}\left(\widetilde P_{\Psi_\lambda},\widetilde P_{\Phi_\lambda}\right)
    =0,
\end{align}
which completes the proof.
\end{proof}

\begin{center}
{\it Remark on conversions between energy eigenstates}
\end{center}

So far, we have considered input and output states whose symmetry subgroups coincide but are not equal to $\mathbb{R}$: $\mathrm{Sym}(\psi) = \mathrm{Sym}(\phi)\subsetneq \mathbb{R}$. However, the conclusion of Proposition~\ref{supp:prop:input-output-energy-distributions} remains valid in the case of $\mathrm{Sym}(\psi) = \mathrm{Sym}(\phi)= \mathbb{R}$, in which both states are energy eigenstates. Here we briefly explain the construction for this case. Since energy eigenstates have zero energy variance, the auxiliary bath $B'$, which was used to absorb the difference in energy variance, is not needed, and it suffices to use bath $B$.

For a fixed typical state $\lambda\in\mathcal{T}_\ell$ of bath $B$, we compare the energy distributions of $\ket{\Psi_\lambda} := \ket{\psi}^{\otimes n}\ket{\lambda}$ and $\ket{\Phi_\lambda} := \ket{\phi}^{\otimes m_n}\ket{t(\lambda)}$.
We write the energy eigenvalues of $\ket{\psi}$ and $\ket{\phi}$ as $E_\psi$ and $E_\phi$, respectively. We first construct the output occupation vector $\bm{t}$ of the output state $\ket{t(\lambda)} \in\mathcal{H}_{\widetilde{B}}$. We denote the occupation vector of the input state $\ket{\lambda}$ by $\bm{s}$. We set $\bm{t} = \bm{s} + \bm{b}^{(n)}$ such that $\bm{b}^{(n)}\in\mathbb{Z}^d$, independent of $\lambda$, satisfies 
\begin{align}
    \sum_{i=1}^d b_i^{(n)} = n-m_n
\end{align}
and 
\begin{align}\label{supp:eq:eigenstate-Eb}
    \bm{E}\cdot\bm{b}^{(n)} = nE_\psi - m_nE_\phi.
\end{align}
By the same argument used in the proof of Lemma~\ref{supp:lem:output-occupation-vector}, we conclude that there exists such $\bm{b}^{(n)}$, and, for sufficiently large $n$, $\bm{t} = \bm{s} + \bm{b}^{(n)}$ is a valid occupation vector of bath $\widetilde{B}$. Eq.~\eqref{supp:eq:eigenstate-Eb} ensures that $\ket{\Psi_\lambda}$ and $\ket{\Phi_\lambda}$ are energy eigenstates with exactly the same energy. Thus, the total variation distance between $P_{\Psi_\lambda}$ and $P_{\Phi_\lambda}$ is zero, and hence we conclude that Proposition~\ref{supp:prop:input-output-energy-distributions} still holds in the case of $\mathrm{Sym}(\psi) = \mathrm{Sym}(\phi)= \mathbb{R}$.

The remainder of the argument applies, with minor modifications, to any pair of states $\psi$ and $\phi$ with the same time-translation symmetry subgroup, including energy eigenstates. For clarity, however, we continue to use the notation appropriate for states with $\mathrm{Sym}(\psi) = \mathrm{Sym}(\phi)\subsetneq \mathbb{R}$, and comment on the necessary modifications for energy eigenstates at the relevant points.

\subsubsection{Conversion for all typical bath states}
\label{supp:subsec:all-typical}

In the previous section, we showed that the total variation distance between the energy distributions of the input and target states vanishes uniformly over all typical bath states $\lambda\in\mathcal{T}_\ell$. In order to construct the desired thermal operation, we need a single energy-conserving unitary that converts $\ket{\Psi_\lambda}$ to $\ket{\Phi_\lambda}$ for all typical bath states $\lambda\in\mathcal{T}_\ell$, which we construct in this section. We first show that, for every typical occupation vector $\bm{s}$, the dimension of the output energy eigenspace is sufficiently large so that we can allocate orthogonal output states to all input states. Then we construct a single energy-conserving unitary $V$ that implements the transformation $\ket{\Psi_\lambda} \to \ket{\Phi_\lambda}$ with a vanishing error for all $\lambda\in\mathcal{T}_\ell$.

\begin{center}
    \textit{Sufficient output degeneracy for orthogonal state assignment}
\end{center}

To construct a single energy-conserving unitary converting all typical input states appropriately, we need to ensure that $\{\ket{\Phi_\lambda}\}_{\lambda\in\mathcal{T}_\ell}$ are orthogonal to each other. To this end, we show that for every typical occupation vector $\bm{s}$ of the input bath $B$, there are sufficiently many output states $\ket{t(\lambda)}$ with occupation vector $\bm{t}=\bm{s}+\bm{b}^{(n)}$, and thus we can assign a distinct output state $\ket{t(\lambda)}$ of $\widetilde{B}$ with an appropriate occupation vector for each typical state $\ket{\lambda}$ of bath $B$.

\begin{proposition}[Orthogonal assignment of output bath states]
\label{supp:prop:orthogonal-output-assignment}
    For all sufficiently large $n$, one can choose, for every typical bath state $\lambda=(\bm s,\alpha)\in\mathcal T_\ell$, an energy eigenstate $\ket{t(\lambda)}\in\mathcal H_{\widetilde B}$ with $t(\lambda)=(\bm t,\alpha')=(\bm s+\bm b^{(n)},\alpha')$ such that the map
    \begin{align}
        \lambda\longmapsto t(\lambda)
    \end{align}
    is injective. Consequently, the output bath states can be chosen to satisfy
    \begin{align}
        \braket{t(\lambda)}{t(\lambda')}=\delta_{\lambda,\lambda'}
    \end{align}
    for all $\lambda,\lambda'\in\mathcal T_\ell$.
\end{proposition}

\begin{proof}
    An input subspace with occupation vector $\bm{s}$ has $M_\mathrm{in}(\bm s)$ orthogonal states labeled by $\alpha = 1, 2, \ldots , M_\mathrm{in}(\bm s)$ with
    \begin{align}
        M_\mathrm{in}(\bm s):=\frac{\ell!}{s_1!s_2!\cdots s_d!}.
    \end{align}
    Similarly, there are $M_\mathrm{out}(\bm t)$ orthogonal states with occupation vector $\bm t$ labeled by $\alpha' = 1, 2, \ldots , M_\mathrm{out}(\bm t)$ with
    \begin{align}
        M_\mathrm{out}(\bm t)
        :=
        \frac{\ell'!}{t_1!t_2!\cdots t_d!}
    \end{align}
    Using Stirling's formula, the logarithms of these two dimensions are computed as
    \begin{align}
        \log M_\mathrm{in}(\bm s) = \ell H\left(\frac{\bm{s}}{\ell}\right) + O(\log \ell),
        \qquad
        \log M_\mathrm{out}(\bm t) = \ell' H\left(\frac{\bm t}{\ell'}\right) + O(\log \ell').
    \end{align}
    Here $H(\bm{p})$ denotes the Shannon entropy of a probability vector $\bm p = (p_1,\ldots,p_d)$:
    \begin{align}
        H(\bm{p}) := - \sum_{i=1}^d p_i\log p_i.
    \end{align}
    Since $\bm{s}$ is an occupation vector of a typical state, its difference from $\ell \bm{g}$ is bounded as
    \begin{align}
        \|\bm{s}  -\ell \bm{g}\|_\infty = O(\sqrt{\ell}\log \ell) =  O(n^{3/4}\log n),
    \end{align}
    where $\bm{g}$ is the Gibbs distribution of $H_S$. Recall that $\bm b^{(n)}$ satisfies
    \begin{align}\label{supp:eq:Ebapprox}
        \bm E \cdot \bm b^{(n)} &= nE_\psi^\mathrm{min} - m_n E_\phi^\mathrm{min} + a_n E_\tau\nonumber\\
        &=nE_\psi -  m_nE_\phi - E_\tau \Delta_n + O(1)\nonumber\\
        &=nE_\psi -  m_nE_\phi + O(n^{3/4}),
    \end{align}
    where we used $\Delta_n = \lfloor n^{3/4}\rfloor$. If $\mathrm{Sym}(\psi)=\mathrm{Sym}(\phi)=\mathbb{R}$, the last line of Eq.~\eqref{supp:eq:Ebapprox} follows directly from Eq.~\eqref{supp:eq:eigenstate-Eb}, with the $O(n^{3/4})$ term replaced by zero. Expanding  $\log M_\mathrm{out}(\bm t) - \log M_\mathrm{in}(\bm s)$ to leading order in $n$, we obtain
    \begin{align}
        \log M_\mathrm{out}(\bm t) - \log M_\mathrm{in}(\bm s)
        &=
        (n+\ell-m_n)H\left(\frac{\bm s + \bm b^{(n)}}{n+\ell-m_n}\right) - \ell H\left(\frac{\bm s}{\ell}\right) + O(\log \ell)\nonumber\\
        &=
        (n-m_n)H(\bm g) + \beta \bm E \cdot \left(\bm b^{(n)} - (n-m_n)\bm g \right) + O(n^{3/4}\log n)\nonumber\\
        &=
        n\left( \beta E_\psi - \beta \bm E\cdot \bm g + H(\bm g)\right) - m_n\left( \beta E_\phi - \beta \bm E\cdot \bm g + H(\bm g)\right) + O(n^{3/4}\log n)\nonumber\\
        &=
        nD(\psi\|\gamma_S) - m_n D(\phi\|\gamma_S) + O(n^{3/4}\log n).\label{supp:eq:degeneracy}
    \end{align}
    Since 
    \begin{align}
        \lim_{n\to\infty}\frac{m_n}{n} = R < R^*(\psi\to\phi)\le\frac{D(\psi\|\gamma_S)}{D(\phi\|\gamma_S)},
    \end{align}
    we conclude that
    \begin{align}
        M_\mathrm{out}(\bm t) > M_\mathrm{in}(\bm s)
    \end{align}
    for sufficiently large $n$. Therefore, for each typical state $\lambda$ with occupation vector $\bm{s}$, we can assign mutually orthogonal states with occupation vector $\bm{t}$. Moreover, since $\bm{t} = \bm{s} + \bm{b}^{(n)}$, different occupation vectors $\bm{s}\neq \bm{s}'$ correspond to different output occupation vectors $\bm{t}\neq\bm{t}'$. Hence, we can take an injective map $\lambda\mapsto t(\lambda)$, which ensures $\braket{t(\lambda)}{t(\lambda')} = \delta_{\lambda,\lambda'}$ for all typical bath states $\lambda,\lambda'\in \mathcal{T}_\ell.$
\end{proof}

\begin{center}
    \textit{Construction of a single energy-conserving unitary}
\end{center}

Here we construct a single energy-conserving unitary which maps $\ket{\Psi_\lambda}$ to a state close to $\ket{\Phi_\lambda}$ for all typical states of bath $B$.

\begin{proposition}[Existence of a single energy-conserving unitary for all typical states]\label{supp:prop:conveersion-for-all}
    There exists an energy-conserving unitary $V$ such that
    \begin{align}\label{supp:eq:single-unitary-V}
        \lim_{n\to\infty}\sup_{\lambda\in\mathcal{T}_\ell}\left\|V\ket{\Psi_\lambda} - \ket{\Phi_\lambda}\right\| = 0.
    \end{align}
\end{proposition}
\begin{proof}
    If $\mathrm{Sym}(\psi)=\mathrm{Sym}(\phi)=\mathbb{R}$, all factors associated with $B'$ are omitted throughout this proof. We write the decomposition of the total Hilbert space into energy subspaces as
    \begin{align}
        \mathcal{H}_\mathrm{tot} 
        =
        \mathcal{H}_S^{\otimes (n+\ell)}\otimes \mathcal{H}_{B'}
        =
        \bigoplus_E \mathcal{H}_E,
    \end{align}
    where $\mathcal{H}_E$ is a subspace of $\mathcal{H}_\mathrm{tot}$ with energy $E$. Let $\Pi_E$ denote the projector onto $\mathcal{H}_E$. We define the weights of $\ket{\Psi_\lambda}$ and $\ket{\Phi_\lambda}$ in each energy sector by
    \begin{align}
        p_{\lambda, E} := \|\Pi_E \ket{\Psi_\lambda}\|^2, \qquad q_{\lambda, E} := \|\Pi_E \ket{\Phi_\lambda}\|^2.
    \end{align}
    For a fixed $\lambda\in\mathcal{T}_\ell$, we define the set of energies in which both $\ket{\Psi_\lambda}$ and $\ket{\Phi_\lambda}$ have nonzero components as
    \begin{align}
        F_\lambda := \{E: p_{\lambda, E}>0\text{ and } q_{\lambda, E}>0\},
    \end{align}
    and define the corresponding set of typical labels by
    \begin{align}
        \Lambda_E := \{\lambda\in\mathcal{T}_\ell: E\in F_\lambda\}.
    \end{align}
    For $E\in F_\lambda$, we define normalized states $\ket{x_{\lambda, E}}, \ket{y_{\lambda, E}}$ by
    \begin{align}
        \ket{x_{\lambda, E}}:= \frac{\Pi_E \ket{\Psi_\lambda}}{\sqrt{p_{\lambda, E}}}, \qquad
        \ket{y_{\lambda, E}}:= \frac{\Pi_E \ket{\Phi_\lambda}}{\sqrt{q_{\lambda, E}}}.
    \end{align}
    Then, each of the two sets $\{\ket{x_{\lambda, E}}\}_{\lambda\in \Lambda_E}$ and $\{\ket{y_{\lambda, E}}\}_{\lambda\in \Lambda_E}$ is  orthonormal. We verify this as follows. We define a projector $P_\lambda:=I_S^{\otimes n}\otimes \ketbra{\lambda}{\lambda}\otimes I_{B'}$, which satisfies $P_\lambda\ket{\Psi_\lambda} = \ket{\Psi_\lambda},$ $[P_\lambda, \Pi_E] = 0,$ and $P_\lambda P_{\lambda'} = \delta_{\lambda, \lambda'}P_\lambda$. Using these properties, we get
    \begin{align}
        \braket{x_{\lambda, E}}{x_{\lambda', E}}=
        \frac{\mel{\Psi_{\lambda}}{\Pi_E}{\Psi_{\lambda'}}}{\sqrt{p_{\lambda, E}\ p_{\lambda', E}}} = 
        \frac{\mel{\Psi_{\lambda}}{P_\lambda\Pi_EP_{{\lambda'}}}{\Psi_{\lambda'}}}{\sqrt{p_{\lambda, E}\ p_{\lambda', E}}} = 
        \frac{\mel{\Psi_{\lambda}}{P_\lambda P_{\lambda'}\Pi_E}{\Psi_{\lambda'}}}{\sqrt{p_{\lambda, E}\ p_{\lambda', E}}} =
        \delta_{\lambda, \lambda'}.
    \end{align}
    For $\{\ket{y_{\lambda, E}}\}_{\lambda\in \Lambda_E}$, we define a projector $Q_\lambda:=I_S^{\otimes m_n}\otimes \ketbra{t(\lambda)}{t(\lambda)}\otimes I_{B'}$. In addition to the obvious relations $Q_\lambda\ket{\Phi_\lambda} = \ket{\Phi_\lambda}$ and $[Q_\lambda, \Pi_E] = 0$, the orthogonality of $\{\ket{t(\lambda)}\}_{\lambda\in\mathcal{T}_\ell}$ (Proposition~\ref{supp:prop:orthogonal-output-assignment}) ensures that $Q_\lambda$ also satisfies $Q_\lambda Q_{\lambda'}=\delta_{\lambda,\lambda'}Q_\lambda$. Thus, by the same calculation as that for $\{\ket{x_{\lambda, E}}\}_{\lambda\in \Lambda_E}$, we conclude that $\braket{y_{\lambda, E}}{y_{\lambda', E}}=\delta_{\lambda,\lambda'}$.
    
    The orthogonality of the two sets $\{\ket{x_{\lambda, E}}\}_{\lambda\in \Lambda_E}$ and $\{\ket{y_{\lambda, E}}\}_{\lambda\in \Lambda_E}$ implies that there exists an isometry $W_E$ satisfying $W_E\ket{x_{\lambda,E}} = \ket{y_{\lambda,E}}$ for all $\lambda\in\Lambda_E$. Extending the isometry $W_E$, we obtain a unitary $V_E$ on $\mathcal{H}_E$ such that 
    \begin{align}
        V_E\ket{x_{\lambda,E}} = \ket{y_{\lambda,E}}
    \end{align}
    for all $\lambda\in\Lambda_E$. We then define an energy-conserving unitary $V$ by
    \begin{align}
        V = \bigoplus_E V_E.
    \end{align}
    It remains to show that the energy-conserving unitary $V$ satisfies Eq.~\eqref{supp:eq:single-unitary-V}. For every $\lambda\in\mathcal{T}_\ell$, we have
    \begin{align}
        \mel{\Phi_\lambda}{V}{\Psi_\lambda}=
        \sum_E
        \mel{\Phi_\lambda}{\Pi_E V_E \Pi_E}{\Psi_\lambda}=
        \sum_{E\in F_\lambda}\sqrt{p_{\lambda,E}q_{\lambda,E}}\ge
        1-d_{\mathrm{TV}}(P_{\Psi_\lambda},P_{\Phi_\lambda}),
    \end{align}
    where we used the Fuchs-van de Graaf inequality. Hence, we have
    \begin{align}\label{supp:eq:V-error-bound}
        \left\|V\ket{\Psi_\lambda}-\ket{\Phi_\lambda}\right\|^2=
        2-2\mel{\Phi_\lambda}{V}{\Psi_\lambda}\le
        2d_{\mathrm{TV}}\left(P_{\Psi_\lambda},P_{\Phi_\lambda}\right).
    \end{align}
    Combining Eq.~\eqref{supp:eq:V-error-bound} with Proposition~\ref{supp:prop:input-output-energy-distributions}, we obtain
    \begin{align}
        \lim_{n\to\infty}\sup_{\lambda\in\mathcal{T}_\ell}\left\|V\ket{\Psi_\lambda}-\ket{\Phi_\lambda}\right\|
        \le
        \lim_{n\to\infty}\sup_{\lambda\in\mathcal{T}_\ell}\sqrt{2d_{\mathrm{TV}}(P_{\Psi_\lambda},P_{\Phi_\lambda})} = 0,
    \end{align}
    which completes the proof.
    
\end{proof}

\subsubsection{Construction of the thermal operation}

We have already constructed the composite auxiliary system $BB'$ and the energy-conserving unitary $V$. We now show that the resulting thermal operation achieves the desired state conversion.

\begin{proposition}\label{supp:prop:desired-thermal-operation}
    Suppose that $\ket{\psi},\ket{\phi}\in\mathcal{H}_S$ have the same time-translation symmetry subgroup: $\mathrm{Sym}(\psi) = \mathrm{Sym}(\phi)$. Then, for any rate $0<R < R^*(\psi\to\phi)$, there exists a sequence of thermal operations $\{\mathcal{E}_n\}$ such that
    \begin{align}
    \lim_{n\to\infty}d_\mathrm{tr}\left(\mathcal{E}_n(\psi^{\otimes n}), \phi^{\otimes \lfloor Rn \rfloor}\right) = 0.
    \end{align}
\end{proposition}

\begin{proof}
    We first consider the case $\mathrm{Sym}(\psi) = \mathrm{Sym}(\phi)\subsetneq\mathbb{R}$. For a given rate $R$ satisfying $0<R<R^*(\psi\to\phi)$, we define the desired thermal operation for each $n$ by
    \begin{align}
        \mathcal{E}_n(\,\cdot\,) = \Tr_{\widetilde{B}B'}\!\left[V(\,\cdot\,\otimes \gamma_B\otimes \gamma_{B'})V^\dagger\right].
    \end{align}
    Below, we demonstrate that this thermal operation indeed realizes the desired conversion: $\psi^{\otimes n}\to \phi^{\otimes \lfloor Rn\rfloor}$.
    Let
    \begin{align}
        \gamma_B = \sum_\lambda p_\lambda\ketbra{\lambda}{\lambda}
    \end{align}
    be the spectral decomposition of $\gamma_B$. Then, the weight of the typical subspace satisfies $p_\mathrm{typ} = \sum_{\lambda\in\mathcal{T}_\ell} p_\lambda$. Proposition~\ref{supp:prop:conveersion-for-all} and the inequality $d_\mathrm{tr}(V\ketbra{\Psi_\lambda}{\Psi_\lambda}V^\dagger, \ketbra{\Phi_\lambda}{\Phi_\lambda}) \le \|V\ket{\Psi_\lambda} - \Phi_\lambda\|$ ensure that there exists a positive sequence $\{\epsilon_n\}$ with $\lim_{n\to\infty} \epsilon_n = 0$ such that
    \begin{align}
    d_\mathrm{tr}\left(V\ketbra{\Psi_\lambda}{\Psi_\lambda}V^\dagger, \ketbra{\Phi_\lambda}{\Phi_\lambda}\right) \le \epsilon_n
    \end{align}
    for all $\lambda\in\mathcal{T}_\ell$.
    We now prove the claim directly:
    \begin{align}
        &d_\mathrm{tr}\left(\mathcal{E}_n(\psi^{\otimes n}), \phi^{\otimes \lfloor Rn \rfloor}\right)\nonumber\\
        &\le
        p_\mathrm{gr} d_\mathrm{tr}\left(\Tr_{\widetilde{B}B'}\!\left[V(\psi^{\otimes n}\otimes \gamma_B\otimes \ketbra{g}{g})V^\dagger\right], \phi^{\otimes \lfloor Rn \rfloor}\right) + (1-p_\mathrm{gr})\nonumber\\
        &\le
        p_\mathrm{gr} \sum_{\lambda\in\mathcal{T}_\ell}p_\lambda d_\mathrm{tr}\left(\Tr_{\widetilde{B}B'}\!\left[V(\psi^{\otimes n}\otimes \ketbra{\lambda}{\lambda}\otimes \ketbra{g}{g})V^\dagger\right], \phi^{\otimes \lfloor Rn \rfloor}\right) +p_\mathrm{gr}(1-p_\mathrm{typ}) + (1-p_\mathrm{gr})\nonumber\\
        &\le
        p_\mathrm{gr} \sum_{\lambda\in\mathcal{T}_\ell}p_\lambda d_\mathrm{tr}\left(\Tr_{\widetilde{B}B'}\!\left[V\ketbra{\Psi_\lambda}{\Psi_\lambda}V^\dagger\right], \Tr_{\widetilde{B}B'}\!\left[\ketbra{\Phi_\lambda}{\Phi_\lambda}\right]\right) +p_\mathrm{gr}(1-p_\mathrm{typ}) + (1-p_\mathrm{gr})\nonumber\\
        &\le
        p_\mathrm{gr} \sum_{\lambda\in\mathcal{T}_\ell}p_\lambda d_\mathrm{tr}\left(V\ketbra{\Psi_\lambda}{\Psi_\lambda}V^\dagger, \ketbra{\Phi_\lambda}{\Phi_\lambda}\right) +p_\mathrm{gr}(1-p_\mathrm{typ}) + (1-p_\mathrm{gr})\nonumber\\
        &\le
        p_\mathrm{gr}\ p_\mathrm{typ}\ \epsilon_n + p_\mathrm{gr}(1-p_\mathrm{typ}) + (1-p_\mathrm{gr}),
    \end{align}
    where $p_\mathrm{gr}$ is the Gibbs weight of the ground state of bath $B'$ defined in Eq.\eqref{supp:eq:pgr-def}. Since $p_\mathrm{typ}\to 1$, $p_\mathrm{gr}\to 1$, and $\epsilon_n \to 0$ as $n\to\infty$, we obtain
    \begin{align}\label{supp:eq:thermaloperation-demo}
        \lim_{n\to\infty} d_\mathrm{tr}\left(\mathcal{E}_n(\psi^{\otimes n}), \phi^{\otimes \lfloor Rn \rfloor}\right) = 0,
    \end{align}
    In the case of $\mathrm{Sym}(\psi) = \mathrm{Sym}(\phi)=\mathbb{R}$, the construction in Proposition~\ref{supp:prop:conveersion-for-all} gives $V\ket{\Psi_\lambda} = \ket{\Phi_\lambda}$ exactly for all $\lambda\in\mathcal{T}_\ell$. Therefore,
    defining the desired thermal operation by
    \begin{align}
        \mathcal{E}_n(\,\cdot\,) = \Tr_{\widetilde{B}}\!\left[V(\,\cdot\,\otimes \gamma_B)V^\dagger\right],
    \end{align}
    we have
    \begin{align}
        d_\mathrm{tr}\left(\mathcal{E}_n(\psi^{\otimes n}), \phi^{\otimes \lfloor Rn \rfloor}\right)
        &\le
        \sum_{\lambda\in\mathcal{T}_\ell}p_\lambda d_\mathrm{tr}\left(\Tr_{\widetilde{B}}\!\left[V\ketbra{\Psi_\lambda}{\Psi_\lambda}V^\dagger\right], \Tr_{\widetilde{B}}\!\left[\ketbra{\Phi_\lambda}{\Phi_\lambda}\right]\right) +(1-p_\mathrm{typ})\nonumber\\
        &=
        1-p_\mathrm{typ}.
    \end{align}
    Thus, Eq.~\eqref{supp:eq:thermaloperation-demo} also holds for any pair of states $\psi$ and $\phi$ satisfying $\mathrm{Sym}(\psi) = \mathrm{Sym}(\phi)$, which completes the proof.
\end{proof}

\subsection{Achievability when $\mathrm{Sym}(\psi)\subsetneq\mathrm{Sym}(\phi)$}

We consider the case in which $\psi$ and $\phi$ have different symmetry subgroups: $\mathrm{Sym}(\psi)\subsetneq\mathrm{Sym}(\phi)$. In this case, we reduce the conversion problem to one between states with the same symmetry subgroup by retaining one copy of the input state $\psi$ in each output block.

We define two states $\ket{\psi'_M}, \ket{\phi'_M} \in \mathcal{H}_S^{\otimes M}$ for an integer $M\ge 2$ by
\begin{align}
    \ket{\psi'_M} := \ket{\psi}^{\otimes M},
    \qquad
    \ket{\phi'_M} := \ket{\phi}^{\otimes (M-1)}\otimes \ket{\psi}.
\end{align}
Then, $\ket{\psi'_M}$ and $\ket{\phi'_M}$ are two pure states with exactly the same time-translation symmetry subgroup $\mathrm{Sym}(\psi_M')=\mathrm{Sym}(\psi)=\mathrm{Sym}(\psi)\cap\mathrm{Sym}(\phi)=\mathrm{Sym}(\phi_M')$. Moreover, both states are defined on a system with Hamiltonian $H_S^{(M)}$, whose energy-level spacings are integer multiples of the energy unit $E_*$. Thus, Proposition~\ref{supp:prop:desired-thermal-operation} guarantees the existence of a sequence of thermal operations converting  $\ket{\psi'_M}^{\otimes n}$ to $\ket{\phi'_M}^{\otimes \lfloor \widetilde{R}n\rfloor}$ for any conversion rate $\widetilde{R}$ satisfying 
\begin{align}
    \widetilde{R} < R_M^*(\psi\to\phi) :=\min\left\{\frac{D(\psi'_M\|\gamma_S^{\otimes M})}{D(\phi'_M\|\gamma_S^{\otimes M})},
    \frac{V_{H_{S}^{(M)}}(\psi'_M)}{V_{H_{S}^{(M)}}(\phi'_M)}\right\}.
\end{align}

For the input state $\ket{\psi}^{\otimes n}$, we first group the input copies into blocks of size $M$ and rewrite $\psi^{\otimes n}$ as $\psi_M'^{\otimes \lfloor n/M\rfloor}\otimes \psi^{\otimes (n - M\lfloor  n/M\rfloor)}$. We apply the above operation to $\psi_M'^{\otimes \lfloor n/M\rfloor}$ to obtain a state close to $\phi_M'^{\otimes \lfloor \widetilde{R}\lfloor  n/M\rfloor\rfloor}$. In the final step, we discard a single copy of $\psi$ in each output block $\phi_M' = \phi^{\otimes (M-1)}\otimes \psi$ and the remaining input copies $\psi^{\otimes (n - M\lfloor  n/M\rfloor)}$. This protocol achieves every conversion rate $R$ satisfying
\begin{align}
    R<\frac{M-1}{M}R_M^*(\psi\to\phi).
\end{align}
Using the additivity of relative entropy and energy variance, we have
\begin{align}
    \lim_{M\to\infty}\frac{M-1}{M}R_M^*(\psi\to\phi)
    &=
    \lim_{M\to\infty}\frac{M-1}{M}\min\left\{\frac{MD(\psi\|\gamma_S)}{(M-1)D(\phi\|\gamma_S)+D(\psi\|\gamma_S)},\frac{MV_{H_S}(\psi)}{(M-1)V_{H_S}(\phi)+V_{H_S}(\psi)}\right\}\nonumber\\
    &=R^*(\psi\to\phi).
\end{align}
Thus we conclude that, by choosing $M$ sufficiently large, every rate below the claimed optimal rate $R^*(\psi\to\phi)$ is achievable. This completes the proof of achievability in the case of $\mathrm{Sym}(\psi)\subsetneq \mathrm{Sym}(\phi)$.

\section{Proofs of technical lemmas}\label{supp:sec:technical-lemmas}

\subsection{Proof of Lemma~\ref{supp:lem:convolution-contraction}}

We prove Lemma~\ref{supp:lem:convolution-contraction}. By the triangle inequality, we have
\begin{align}
d_{\mathrm{TV}}\left(p_1*q_1,p_2*q_2\right)
&\leq
d_{\mathrm{TV}}\left(p_1*q_1,p_2*q_1\right)
+
d_{\mathrm{TV}}\left(p_2*q_1,p_2*q_2\right).
\label{supp:eq:convolution-tv-triangle}
\end{align}
For the first term in Eq.~\eqref{supp:eq:convolution-tv-triangle}, using the definition of convolution and the triangle
inequality, we have
\begin{align}
d_{\mathrm{TV}}\left(p_1*q_1,p_2*q_1\right)
&=
\frac{1}{2}\sum_{x\in\mathbb Z}\left|\sum_{y\in\mathbb Z}\left(p_1(y)-p_2(y)\right)q_1(x-y)\right|\nonumber\\
&\leq
\frac{1}{2}\sum_{x\in\mathbb Z}\sum_{y\in\mathbb Z}\left|p_1(y)-p_2(y)\right|q_1(x-y)\nonumber\\
&=
\frac{1}{2}\sum_{y\in\mathbb Z}\left|p_1(y)-p_2(y)\right|\sum_{x\in\mathbb Z}q_1(x-y)\nonumber\\
&=
d_{\mathrm{TV}}\left(p_1,p_2\right).
\label{supp:eq:convolution-tv-first-term}
\end{align}
Similarly, we obtain
\begin{align}
d_{\mathrm{TV}}\left(p_2*q_1,p_2*q_2\right)
&\leq
d_{\mathrm{TV}}\left(q_1,q_2\right).
\label{supp:eq:convolution-tv-second-term}
\end{align}
Combining Eqs.~\eqref{supp:eq:convolution-tv-triangle}--\eqref{supp:eq:convolution-tv-second-term}, we obtain  Eq.~\eqref{supp:eq:convolution-tv-bound}.

\subsection{Proof of Lemma~\ref{supp:lem:shifted-poisson}}

We prove Lemma~\ref{supp:lem:shifted-poisson}. Let $P_n$ and $Q_n$ be random variables whose distributions are given by
\begin{align}
    P_n\sim P(\lambda_n),\qquad
    Q_n\sim P(\nu_n),
\end{align}
where $P(v)$ denotes the Poisson distribution with parameter $v$. By the definition of the shifted Poisson distribution, we have
\begin{align}
    SP(s_n,\lambda_n)=\mathcal{L}(s_n+P_n),
    \qquad
    SP(t_n,\nu_n)
    =
    \mathcal L(t_n+Q_n),
\end{align}
where $\mathcal L(X)$ is the probability distribution of a random variable $X$. Set
\begin{align}
    k_n:=t_n-s_n.
\end{align}
By assumption, $k_n\in\mathbb Z$ and $|k_n|=O(1)$. Translation invariance and the triangle inequality for the total variation distance give
\begin{align}
    d_{\mathrm{TV}}\left(SP(s_n,\lambda_n),SP(t_n,\nu_n)\right)\nonumber
    &=
    d_{\mathrm{TV}}\left(\mathcal L(P_n),\mathcal L(Q_n+k_n)\right)\nonumber\\
    &\leq
    d_{\mathrm{TV}}\left(\mathcal L(P_n),\mathcal L(Q_n)\right)
    +
    d_{\mathrm{TV}}\left(\mathcal L(Q_n),\mathcal L(Q_n+k_n)\right).
    \label{supp:eq:shifted-poisson-two-terms}
\end{align}
We first estimate the first term in Eq.~\eqref{supp:eq:shifted-poisson-two-terms}. For Poisson distributions, the bound 
\begin{align}
    d_{\mathrm{TV}}\left(P(\lambda_n),P(\nu_n)\right)\le
    \frac{|\lambda_n-\nu_n|}{\sqrt{\min\{\lambda_n,\nu_n\}}}
\end{align}
follows from Ref.~\cite{supp:Adell-2006}. Therefore, the assumptions $|\lambda_n-\nu_n|=O(1)$ and $\lambda_n,\nu_n=\Theta(n)$ imply that
\begin{align}
    d_{\mathrm{TV}}
    \left(\mathcal L(P_n),\mathcal L(Q_n)\right)
    =
    O(n^{-1/2}).
    \label{supp:eq:poisson-parameter-tv-bound}
\end{align}
It remains to estimate the second term in Eq.~\eqref{supp:eq:shifted-poisson-two-terms}. Let
\begin{align}
    q_n(j):=\Pr[Q_n=j],\qquad
    j\in\mathbb Z,
\end{align}
where we set $q_n(j)=0$ for $j<0$. We first consider the case
$k_n\ge0$. Then
\begin{align}
    d_{\mathrm{TV}}
    \left(\mathcal L(Q_n),\mathcal L(Q_n+k_n)\right)
    &=
    \frac{1}{2}\sum_{j\in\mathbb Z}\left|q_n(j)-q_n(j-k_n)\right|\nonumber\\
    &\le
    \frac{1}{2}\sum_{j\in\mathbb Z}\sum_{r=0}^{k_n-1}\left|q_n(j-r)-q_n(j-r-1)\right|\nonumber\\
    &=
    \frac{k_n}{2}\sum_{j\in\mathbb Z}\left|q_n(j)-q_n(j-1)\right|.
    \label{supp:eq:bounded-poisson-shift}
\end{align}
Set $q_n(m) = \max_{j\in\mathbb Z}q_n(j)$. Since
\begin{align}
    q_n(0) \le q_n(1) \le \cdots \le q_n(m-1)\le q_n(m) \ge q_n(m+1) \ge\cdots
\end{align}
we have
\begin{align}
    \sum_{j\in\mathbb Z}\left|q_n(j)-q_n(j-1)\right|=2q_n(m).
\end{align}
Moreover, Stirling's formula gives
\begin{align}
    \max_{j\in\mathbb{Z}}q_n(j)=O(\nu_n^{-1/2}).
\end{align}
Since $k_n=O(1)$ and $\nu_n=\Theta(n)$, we obtain
\begin{align}
    d_{\mathrm{TV}}\left(\mathcal L(Q_n),\mathcal L(Q_n+k_n)\right)=O(n^{-1/2}).
    \label{supp:eq:poisson-shift-tv-bound}
\end{align}
The same argument applies when $k_n<0$, with $|k_n|$ in place of $k_n$. Combining Eqs.~\eqref{supp:eq:shifted-poisson-two-terms}, \eqref{supp:eq:poisson-parameter-tv-bound}, and \eqref{supp:eq:poisson-shift-tv-bound}, we conclude that
\begin{align}
    \lim_{n\to\infty}d_{\mathrm{TV}}\left(SP(s_n,\lambda_n),SP(t_n,\nu_n)\right)
    =0
\end{align}
which completes the proof.

\subsection{Proof of Lemma~\ref{supp:lem:input-output-ideal}}
We define $\gamma_{\psi,n}, \gamma_{\phi,n},$ and $\gamma_{\eta,n}$ by
\begin{align}
    \gamma_{\psi,n}
    &:=
    n\mu_\psi-n\sigma_\psi^2-\left\lfloor n\mu_\psi-
    n\sigma_\psi^2\right\rfloor,\\
    \gamma_{\phi,n}
    &:=
    m_n\mu_\phi-m_n\sigma_\phi^2-\left\lfloor m_n\mu_\phi-m_n\sigma_\phi^2\right\rfloor,\\
    \gamma_{\eta,n}
    &:=
    \Delta_n-\Sigma_n^2-\left\lfloor\Delta_n-\Sigma_n^2\right\rfloor.
\end{align}
By definition, $0\le\gamma_{\psi,n},\gamma_{\phi,n},\gamma_{\eta,n}<1$. Using the representation of a translated Poisson distribution as a shifted Poisson distribution (see Eq.~\eqref{supp:eq:tp-as-sp-preliminaries}), the input distribution $Q_n^{\mathrm{in}}$ can be written as
\begin{align}
    Q_n^{\mathrm{in}}=SP\left(s_n^{\mathrm{in}},v_n^{\mathrm{in}}\right),
\end{align}
where
\begin{align}
    s_n^{\mathrm{in}}:=\left\lfloor n\mu_\psi-n\sigma_\psi^2\right\rfloor,\qquad
    v_n^{\mathrm{in}}:=n\sigma_\psi^2+\gamma_{\psi,n}.
\end{align}
Similarly, since the convolution of shifted Poisson distributions is again a shifted Poisson distribution (see Eq.~\eqref{supp:eq:sp-convolution-preliminaries}), the output distribution can be written as
\begin{align}
    Q_n^{\mathrm{out}}=SP\left(s_n^{\mathrm{out}},v_n^{\mathrm{out}}\right),
\end{align}
where
\begin{align}
    s_n^{\mathrm{out}}:=a_n+\left\lfloor m_n\mu_\phi-m_n\sigma_\phi^2\right\rfloor+\left\lfloor\Delta_n-\Sigma_n^2\right\rfloor,\qquad
    v_n^{\mathrm{out}}:=m_n\sigma_\phi^2+\Sigma_n^2+\gamma_{\phi,n}+\gamma_{\eta,n}.
\end{align}
Using the relation $\Sigma_n^2=n\sigma_\psi^2-m_n\sigma_\phi^2$, we have
\begin{align}
    v_n^{\mathrm{out}}-v_n^{\mathrm{in}}
    &=
    \gamma_{\phi,n}+\gamma_{\eta,n}-\gamma_{\psi,n},
\end{align}
and hence
\begin{align}\label{supp:eq:v-difference}
    \left|v_n^{\mathrm{out}}-v_n^{\mathrm{in}}\right|
    \le 2 = O(1).
\end{align}
Moreover, since $\sigma_\psi^2>0$,
\begin{align}\label{supp:eq:v-order}
    v_n^{\mathrm{in}}=\Theta(n),\qquad
    v_n^{\mathrm{out}}=\Theta(n).
\end{align}
The means of the two shifted Poisson distributions are
\begin{align}\label{supp:eq:sp-mean}
    s_n^{\mathrm{in}}+v_n^{\mathrm{in}}=n\mu_\psi,\qquad
    s_n^{\mathrm{out}}+v_n^{\mathrm{out}}=a_n+m_n\mu_\phi+\Delta_n.
\end{align}
Combining Eq.~\eqref{supp:eq:sp-mean} with Eqs.~\eqref{supp:eq:an-mean-condition} and~\eqref{supp:eq:v-difference}, we obtain
\begin{align}
    \left|s_n^{\mathrm{in}}-s_n^{\mathrm{out}}\right|
    &=
    \left|n\mu_\psi-\left(a_n+m_n\mu_\phi+\Delta_n\right)-v_n^{\mathrm{in}}+v_n^{\mathrm{out}}\right|\nonumber\\
    &\le
    \left|n\mu_\psi-\left(a_n+m_n\mu_\phi+\Delta_n\right)\right|+\left|v_n^{\mathrm{in}}-v_n^{\mathrm{out}}\right|\nonumber\\
    &=
    O(1).\label{supp:eq:s-difference}
\end{align}
Eqs.~\eqref{supp:eq:v-difference},~\eqref{supp:eq:v-order}, and~\eqref{supp:eq:s-difference} ensure that all assumptions of Lemma~\ref{supp:lem:shifted-poisson} are satisfied for $Q_n^\mathrm{in}$ and $Q_n^\mathrm{out}$. We therefore obtain
\begin{align}
    \lim_{n\to\infty}d_{\mathrm{TV}}\left(Q_n^{\mathrm{in}},Q_n^{\mathrm{out}}\right)
    =
    \lim_{n\to\infty}d_{\mathrm{TV}}\left(SP\left(s_n^{\mathrm{in}},v_n^{\mathrm{in}}\right),SP\left(s_n^{\mathrm{out}},v_n^{\mathrm{out}}\right)\right)
    =0,
\end{align}
which completes the proof.

\begingroup
\renewcommand{\addcontentsline}[3]{}

\endgroup

\end{document}